\documentclass{emulateapj}
\usepackage{graphicx}
\usepackage{psfig}
\usepackage{longtable}

\newcommand{\RA}[3]{{#1}^{{\rm h}}{#2}^{{\rm m}}{#3}^{{\rm s}}}
\newcommand{\Dec}[3]{{#1}^{\circ}{#2}'{#3}''}
\newcommand{\E}[1]{\times 10^{#1}}
\newcommand{\twCO}{$^{12}$CO}  \newcommand{\thCO}{$^{13}$CO}

\newcommand{\VLSR}{V_{\rm LSR}}

\begin{document}

\title{
MOLECULAR ENVIRONMENT OF THE SUPERNOVA REMNANT IC~443:
DISCOVERY OF THE MOLECULAR SHELLS SURROINDING THE REMNANT
}

\shorttitle{Molecular environment of SNR IC~443}

\author{
Yang Su\altaffilmark{1,2}, Min Fang\altaffilmark{1,2}, 
Ji Yang\altaffilmark{1,2}, Ping Zhou\altaffilmark{3}, 
and Yang Chen\altaffilmark{3}
       }

\affil{
$^1$ Purple Mountain Observatory, Chinese Academy of
Sciences, Nanjing 210008, China \\
$^2$ Key Laboratory of Radio Astronomy, Chinese Academy of
Sciences, Nanjing 210008, China \\
$^3$ Department of Astronomy, Nanjing University, Nanjing 210093, China\\
      }

\begin{abstract}
We have carried out \twCO, \thCO, and C$^{18}$O observations toward 
the mixed morphology supernova remnant (SNR) IC~443. The observations cover a 
1\fdg5$\times$1\fdg5 area and allow us 
to investigate the overall molecular environment of the remnant. Some 
northern and northeastern
partial shell structure of CO gas is around the
remnant. One of the partial shells, about 5$'$ extending 
beyond the northeastern border of the remnant's
bright radio shell, seems to just confine the faint radio halo.
On the other hand, some faint
CO clumps can be discerned along the eastern boundary of the faint
remnant's radio halo. Connecting the eastern CO clumps, the
northeastern partial shell structures, and the northern CO partial
shell, we can see that a half molecular ring structure appears to 
surround the remnant. The LSR velocity of the half-ring structure
is in the range of $-5$~km~s$^{-1}$ to $-2$~km~s$^{-1}$, which
is consistent with that of the $-4$~km~s$^{-1}$ molecular clouds. 
We suggest that the half-ring structure
of the CO emission at $\VLSR\sim-4$~km~s$^{-1}$ is associated with
the SNR. The structures are possibly swept up by 
the stellar winds of SNR IC~443's massive progenitor. 
Based on the $Wide$-$field$ $Infrared$ $Survey$ $Explorer$ and 
the Two Micron All Sky Survey near-IR
database, 62 young stellar object (YSO) candidates are selected
within the radio halo of the remnant. These YSO candidates
concentrated along the boundary of the remnant's bright radio shell
are likely to be triggered by the stellar winds from the massive
progenitor of SNR IC~443.

\end{abstract}


\section{INTRODUCTION}
Stars form from molecular clouds (MCs) and the stellar
radiation and winds heavily sculpt their interstellar
environment. When the massive star dies and explodes as a supernova,
the remnant of it often evolves in such a complex environment.
The distribution of the interstellar medium (ISM) and circumstellar medium
in the vicinity of the supernova remnant (SNR) play
an essential role for our understanding of the morphology
and evolution of SNRs. 
Many SNR--MC interacting systems are investigated 
based on molecular observations, especially \twCO\ lines and their isotopes
(Dickman et al. 1992; Seta et al. 1998).

SNR IC~443 (G189.1+3.0), an SNR--MC interacting system (e.g.,
DeNoyer 1979; Dickman et al. 1992; van Dishoeck et al. 1993), has been 
well studied in multiwavelengths (see Shinn et al.
2011 for a recent review). Because of its large scale ($\sim$0\fdg5) and 
the lack
of the confusion in the direction of the remnant, SNR IC~443 has
become a good laboratory to study the astrophysical phenomena,
e.g., the SNR--MC interaction, the relationship between the
mutiwavelength emission and the remnant's evolution, the star formation
activity near the remnant, and the origin of the cosmic rays
in the vicinity of the SNR.

Braun \& Strom (1986) suggested that SNR IC~443 has evolved inside
the preexisting wind-blown bubble, which likely originated from
the remnant's massive progenitor star. This hypothesis was also
supported by Troja et al. (2006, 2008) based on the analysis of
$XMM$-$Newton$ X-ray observations. According to 21 cm spectral-line
and continuum data, Lee et al. (2008) recently suggested
that SNR IC~443 was into the western rarefied medium.

We performed new millimeter CO observations (covering about 2.25 deg$^2$) 
toward the remnant to investigate the overall ambient MC distribution
environing SNR IC~443.
Based on the new CO observations, 
the half-ring structure in the velocity interval of
$-5$~km~s$^{-1}$ to $-2$~km~s$^{-1}$ is probably associated with SNR IC~443. 
We also explore
the possible physical connection between the half-ring structure
and the stellar winds originating from 
the remnant's massive progenitor star.

The paper is structured as follows. In Section 2, we show the CO
observations and the data reduction. Sections 3 and 4 describe
the main results and the physical discussion, respectively. A brief summary
is given in Section 5.

\begin{figure*}
\centerline{\includegraphics[trim=0mm 0mm 0mm 120mm,scale=0.45,angle=0]{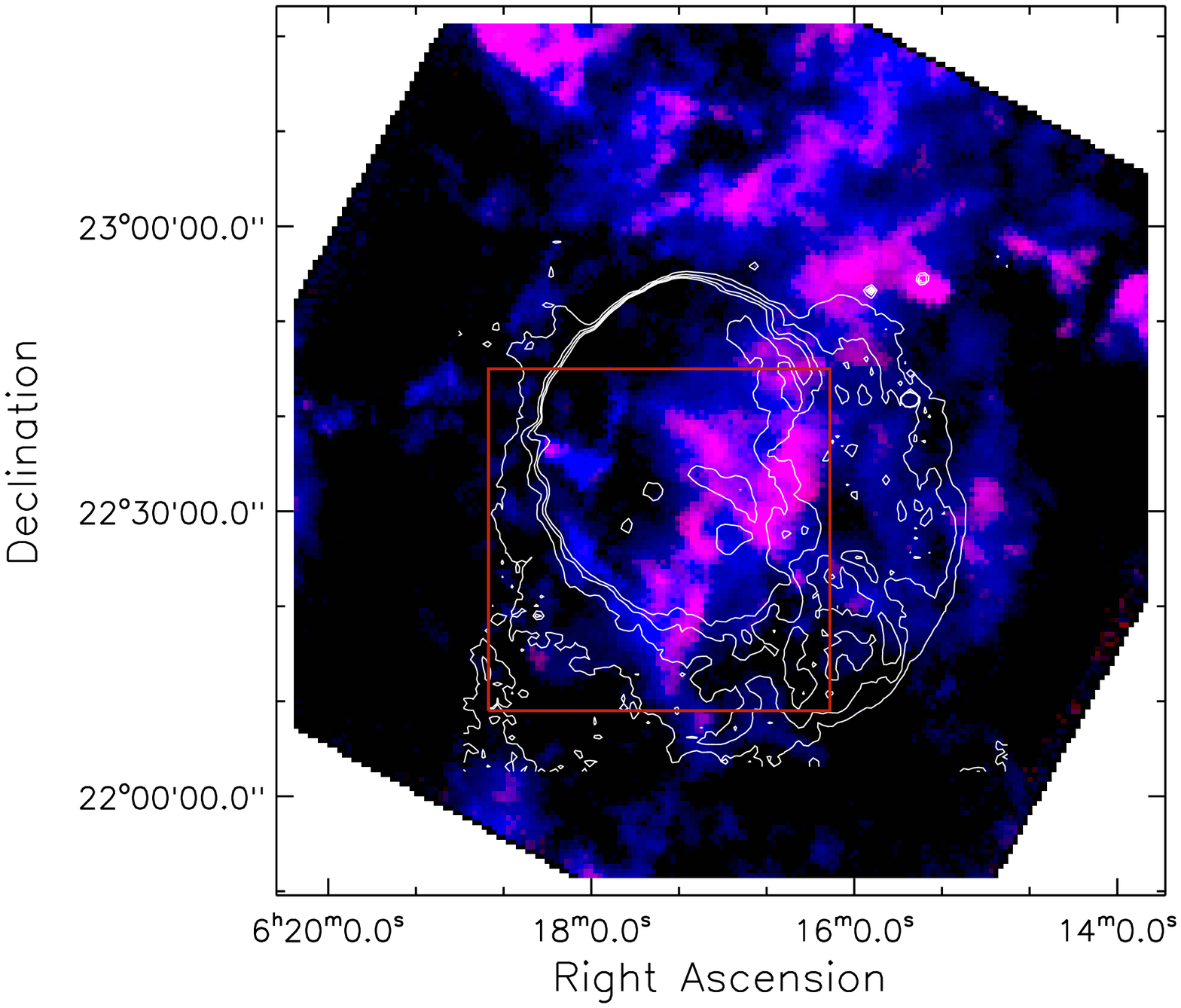}
            \includegraphics[trim=0mm 0mm 0mm 120mm,scale=0.45,angle=0]{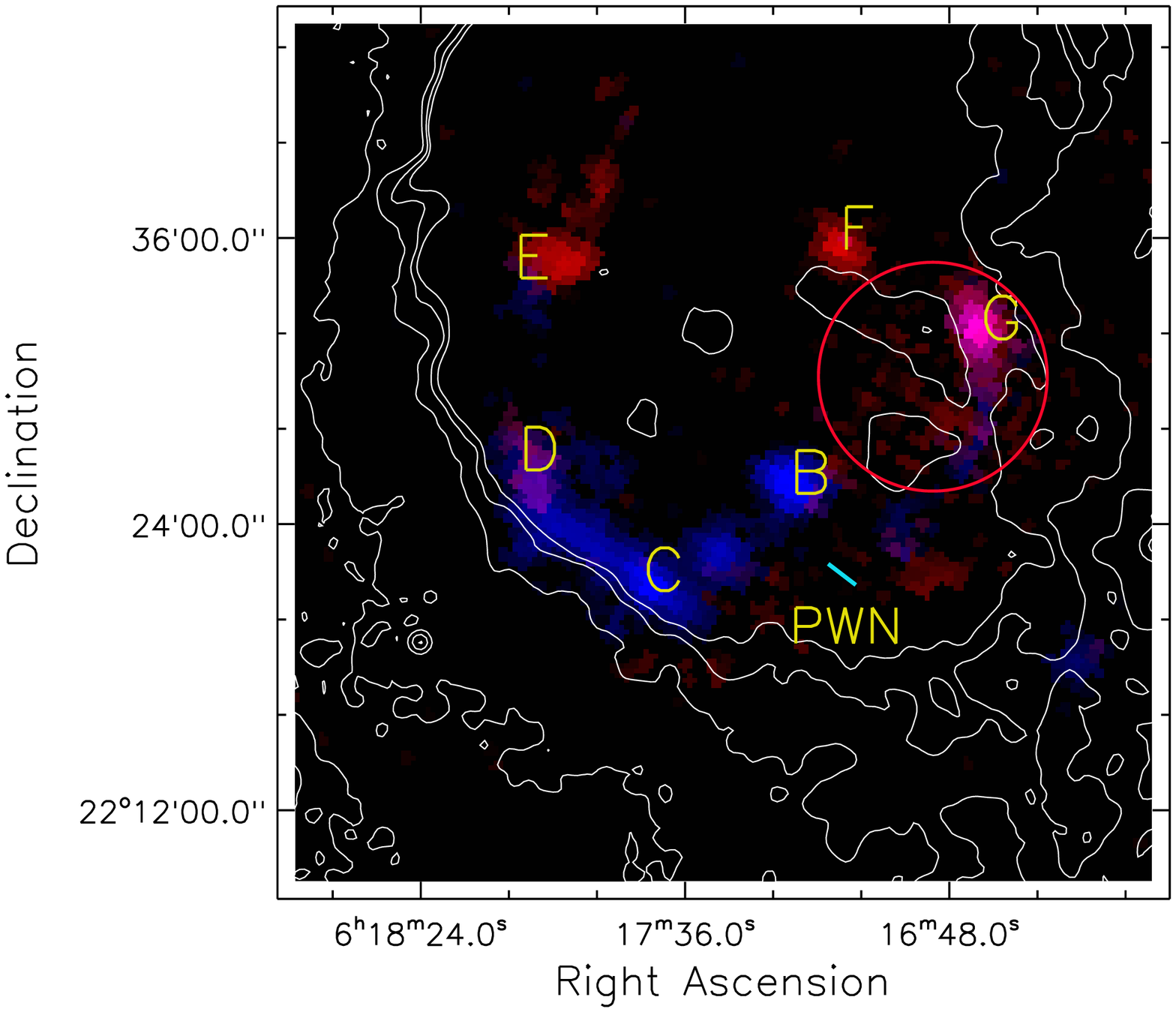}}
	    \caption{
	    Left: \twCO\ ($J$=1--0; blue) and \thCO\ ($J$=1--0; red)
	    intensity maps in the
	    $-10$~km~s$^{-1}$ to 10~km~s$^{-1}$ interval with a square root scale toward
	    SNR IC~443, overlaid with the 1.4~GHz radio continuum emission
	    contours. The red box shows the
	    area where the broadened molecular lines are detected (see right panel).
	    Right: broadened \twCO\ ($J$=1--0) integrated temperature map of
	    SNR IC~443, overlaid
	    with the 1.4~GHz radio continuum emission contours. The blue
	    channel is in the $-50$~km~s$^{-1}$ to $-9$~km~s$^{-1}$ interval and the
	    red is in the
	    0~km~s$^{-1}$ to 4~km~s$^{-1}$ interval. The letters B--G denote the location
	    of the
	    shocked clumps identified by Dickman et al. (1992). The red circle
	    represents the location of the $\gamma$-ray source detected by
	    VERITAS (Acciari et al. 2009), and the cyan line represents the
	    location of the PWN detected by $Chandra$ (Olbert et al. 2001).
	    \label{fig1}}
\end{figure*}

\section{OBSERVATIONS AND DATA REDUCTION}
\subsection{CO Data}

The observations toward SNR IC~443 were made simultaneously in the
\twCO~($J$=1--0) line (at 115.271~GHz), the \thCO~($J$=1--0) line
(110.201~GHz), and the C$^{18}$O~($J$=1--0) line (109.782~GHz) during
2011 November using the 13.7 m millimeter-wavelength telescope
of the Purple Mountain Observatory at Delingha.
It is a part of the Multi-Line Galactic Plane Survey in CO and its 
Isotopic Transitions, also called the Milky Way Imaging Scroll
Painting, for the large-scale 
survey of dense molecular gas along the northern Galactic Plane.
We used a new 3$\times$3 pixel Superconducting Spectroscopic Array 
Receiver as the front end, which was made with 
Supercondcutor-Insulator-Superconductor (SIS) mixers using the 
sideband separating scheme (Shan et al. 2012; Zuo et al. 2011). 
An instantaneous bandwidth of 1~GHz was arranged for the 
backends. Each spectrometer provides 16,384 channels, 
resulting in a spectral resolution of 61 kHz, equivalent to a velocity 
resolution of about 0.16~km~s$^{-1}$ for \twCO\ and 0.17~km~s$^{-1}$ 
for \thCO\ and C$^{18}$O. The half-power beamwidth of the telescope 
is about $54''$ and the pointing accuracy of the telescope is greater 
than $4''$ in the observing epoch.
We used the on-the-fly observing technique
to map the 1\fdg5$\times$1\fdg5 area centered at 
($l$=189\fdg0,$b$=3\fdg0)
with a scan speed of 50$''$~s$^{-1}$ and a step of 15$''$ along the 
Galactic longitude and latitude. The typical system temperature
was around 200--330~K (120--170~K) for \twCO\ (\thCO\ and
C$^{18}$O) for each beam. The mean rms noise level of the brightness
temperature ($T_{\rm R}$) was less than 0.3~K for \twCO\ and
0.2~K for \thCO\ and C$^{18}$O. All of the CO data used in this
study are expressed in brightness temperature, which are
divided by the main beam efficiency $\eta_{\rm mb}=$ 0.46
($T_{\rm R}=T_{\rm A}/(f_{\rm b}\times\eta_{\rm mb})$,
assuming a beam filling factor of $f_{\rm b}\sim$~1).

We have performed the largest \twCO, \thCO, and C$^{18}$O maps
to date for such a moderately fine resolution, with fully sampled
grids and low noise levels, which are essential to
understand the large-scale environment of the remnant.
All of the CO data were reduced using the GILDAS/CLASS
package developed by IRAM.\footnote{http://www.iram.fr/IRAMFR/GILDAS}
Finally, the baseline-corrected spectra were converted to
three-dimensional cube data with a grid spacing of $30''$ and a velocity 
channel separation of 0.2~km~s$^{-1}$ for subsequent analysis.

\begin{deluxetable*}{ccccc}
\tabletypesize{\scriptsize}
\tablecaption{ \thCO\ ($J$=1--0) Spectral Parameters of the Shocked Clumps}
\tablehead{
\colhead{\begin{tabular}{c}
Shocked Clumps \\
Label\\
\end{tabular}} &
\colhead{\begin{tabular}{c}
(R.A., Decl.) \\
(J2000)\\
\end{tabular}} &
\colhead{\begin{tabular}{c}
Extraction Area \\
(arcmin)\\
\end{tabular}} &
\colhead{\begin{tabular}{c}
$\VLSR$(Peak)\\
(km~s$^{-1}$)\\
\end{tabular}} &
\colhead{\begin{tabular}{c}
$T$(Peak)\\
(K)\\
\end{tabular}}
}
\startdata
B & $(\RA{06}{17}{16}.8,\Dec{22}{25}{17})$ & 1$\times$1 & -3.0 & 0.9 \\
C & $(\RA{06}{17}{43}.4,\Dec{22}{21}{13})$ & 1$\times$1.5 &  &  \\
D & $(\RA{06}{18}{06}.0,\Dec{22}{26}{16})$ & 2$\times$2.5 & -7.8 & 0.3 \\
E & $(\RA{06}{18}{06}.9,\Dec{22}{34}{19})$ & 1$\times$1.5 & -4.2 & 0.3 \\
F & $(\RA{06}{17}{07}.8,\Dec{22}{35}{31})$ & 1$\times$1 & -3.4 & 1.5 \\
G & $(\RA{06}{16}{42}.1,\Dec{22}{31}{46})$ & 1$\times$1 & -4.4 & 1.4
\enddata
\tablecomments{
No \thCO\ emission was detected for shocked clump C.}
\end{deluxetable*}

\subsection{Infrared and Radio Data}
The infrared (IR) photometric data used in this work were from
the Two Micron All Sky Survey (2MASS; Skrutskie et al. 2006) and
the survey of the $Wide$-$field$ $Infrared$ $Survey$ $Explorer$
($WISE$, Wright et al. 2010). The 2MASS survey provides the
near-IR photometry in the $JHK_{\rm S}$ bands with 10$\sigma$
limiting magnitudes of $\sim$15.8, 15.1, and 14.3~mag, respectively.
The $WISE$ survey covered the IC~443 region at the wavelengths of 3.4,
4.6, 12, and 22~$\mu$m with spatial resolutions of 6\farcs1, 6\farcs4,
6\farcs5, and 12\farcs0, respectively. The 10$\sigma$ limiting magnitudes are
estimated to be $\sim$15.7, 14.7, 10.4, and 7.1~mag in the $WISE$
[3.4], [4.6], [12], and [22] bands, respectively. In this work,
we use two methods to select the sources. First, we select the sources
with a signal-to-noise ratio (S/N)$>10$ in the first three $WISE$ bands
given the spatial resolution
of the 22~$\mu$m and that strong variable nebulae in the field of IC~443
make the photometry of faint sources very uncertain.
Second, for those sources with a $WISE$ [12] band with an S/N$<10$, we add
the 2MASS $JHK_{\rm S}$ bands to select the young stellar object (YSO)
candidates.
We further reject any source suffering contamination or with confusion
flags (cc\_flag) as ``D", ``H", ``O", or ``P" artifacts
and only selected the point sources with
ext\_flag=0 (Koenig et al. 2012) to reduce the contamination emission by
those extended sources. The detailed method is described by
Koenig et al. (2012, see Appendices A.1--A.4 in their
paper). The results of the selected disk-bearing young stellar
population in the field of view (FOV) of SNR IC~443 are presented
in Section 3.2.

The 1.4 GHz radio continuum emission data were obtained (Lee
et al. 2008). The $Spitzer$ 24~$\mu$m mid-IR observations
were also used for comparison.

\begin{figure}
            \includegraphics[scale=1.6,angle=0]{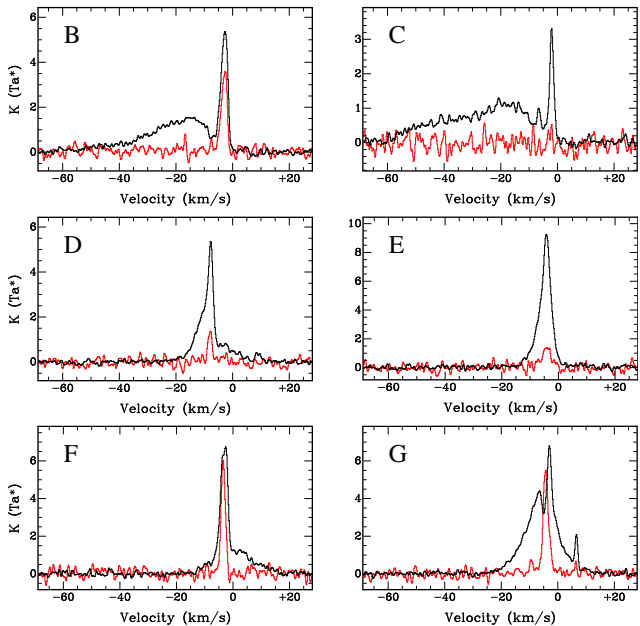}
       	    \caption{
	    \twCO ($J$=1--0; black) and \thCO ($J$=1--0; red, multiplied by a factor
	    of four) spectra of the shocked clumps B--G.
	    \label{fig2}}
\end{figure}

\begin{figure*}
\centerline{\includegraphics[trim=0mm 0mm 0mm 140mm,scale=0.39,angle=0]{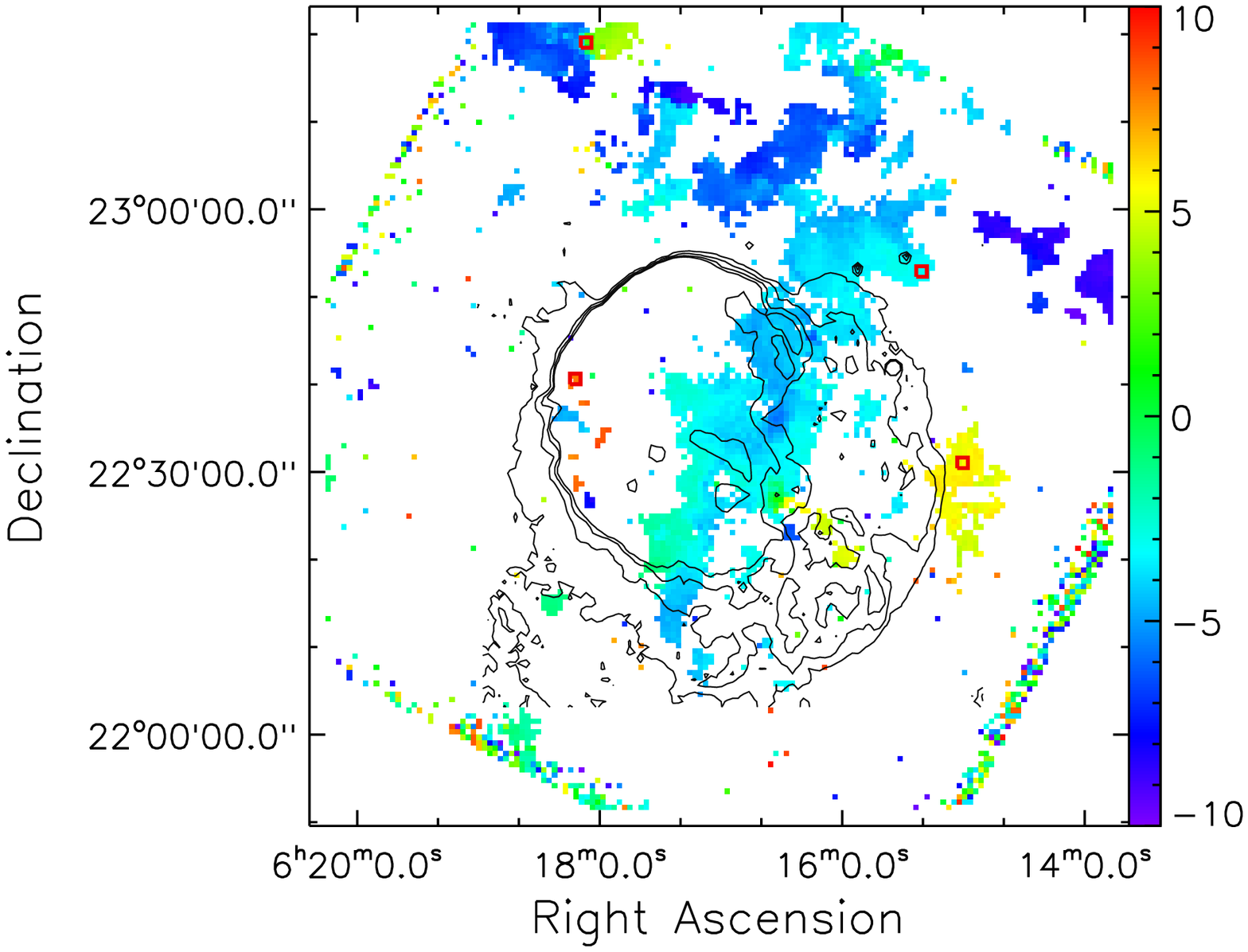}
            \includegraphics[trim=0mm 0mm 0mm 140mm,scale=0.39,angle=0]{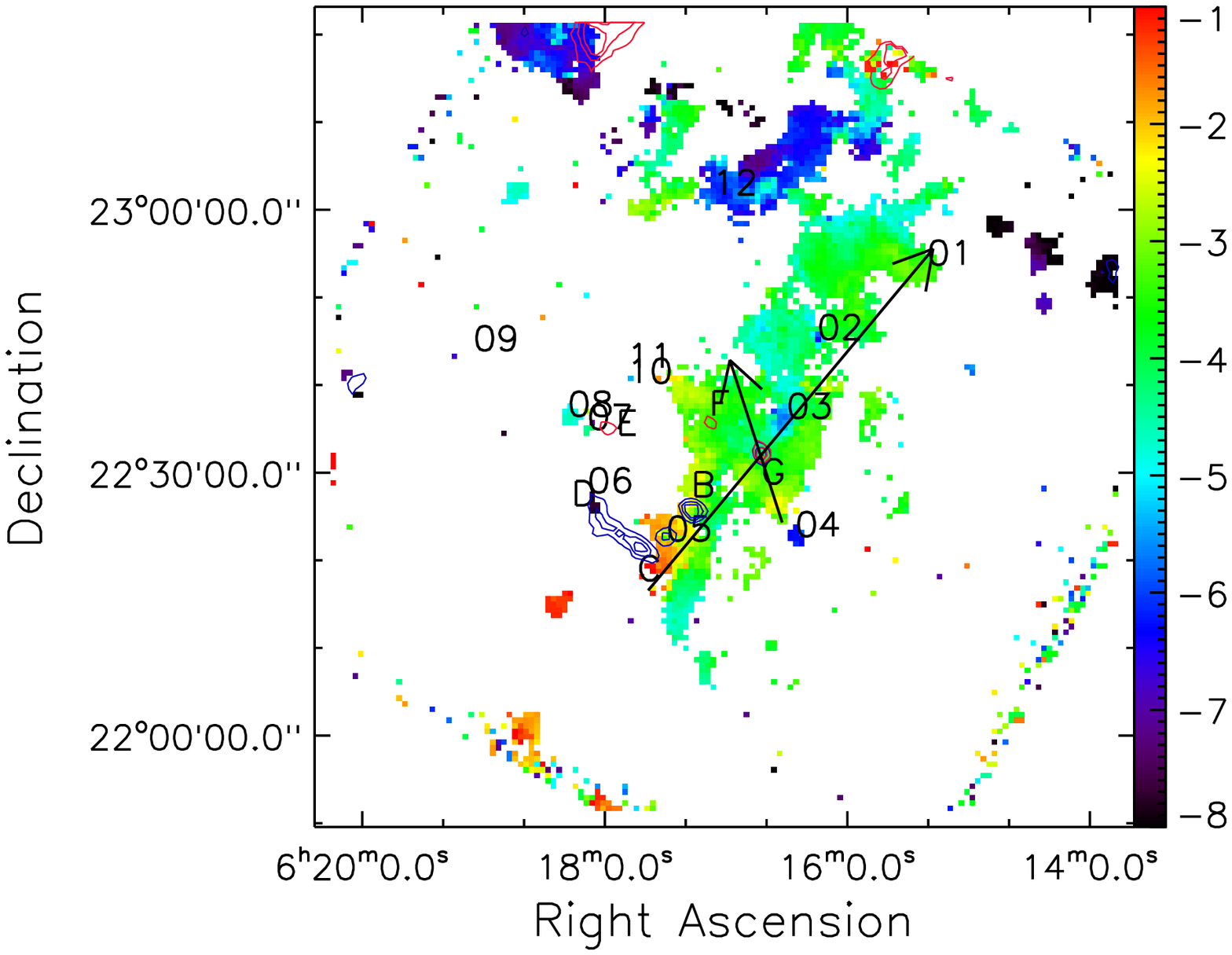}}
	    \caption{Left: the intensity weighted mean velocity (first moment)
	    map of the \thCO\ ($J$=1--0) emission in the interval of $-10$~km~s$^{-1}$
	    to 10~km~s$^{-1}$,
	    overlaid with the 1.4~GHz radio continuum emission contours.
	    The four small red boxes indicate the regions from which the typical spectra
	    for the five MC components are extracted (see Figure~4). Right:
	    the intensity weighted mean velocity (first moment) map of the \thCO\
	    emission in the interval of $-8$~km~s$^{-1}$ to $-1$~km~s$^{-1}$. The numbers
	    01--12 denote the location of the SCs identified by Lee et al. (2012). The
	    letters B--G denote the location of the shocked clumps and the
	    blue and red contours show the blueshifted ($-50$~km~s$^{-1}$ to
	    $-9$~km~s$^{-1}$) and redshifted (0~km~s$^{-1}$ to 4~km~s$^{-1}$) components,
	    respectively (see Figure~1, right panel).
	    \label{fig3}}
\end{figure*}

\begin{figure}
	    \includegraphics[scale=0.65,angle=0]{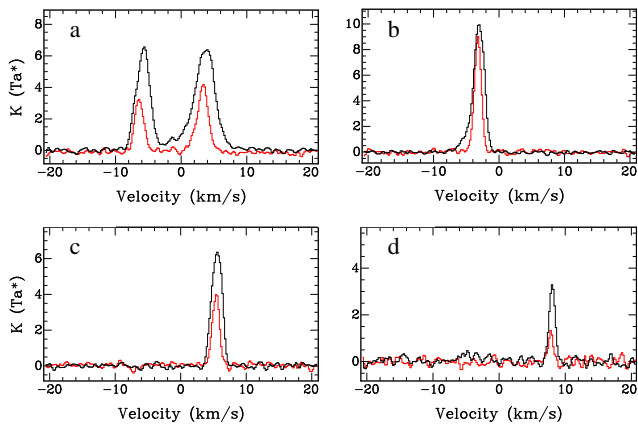}
	    \caption{
	    Typical \twCO ($J$=1--0; black) and \thCO ($J$=1--0;
	    red, multiplied by a factor of two) spectra of the different MC
	    components showing peaks at (a) $-7$~km~s$^{-1}$ and 3~km~s$^{-1}$, (b)
	    $-3$~km~s$^{-1}$, (c) 5~km~s$^{-1}$, and (d) 8~km~s$^{-1}$, respectively.
	    The spectra extraction regions are indicated by the small red box in Figure~3.
	    \label{fig4}}
\end{figure}

\begin{figure}
	    \includegraphics[scale=1.5,angle=0]{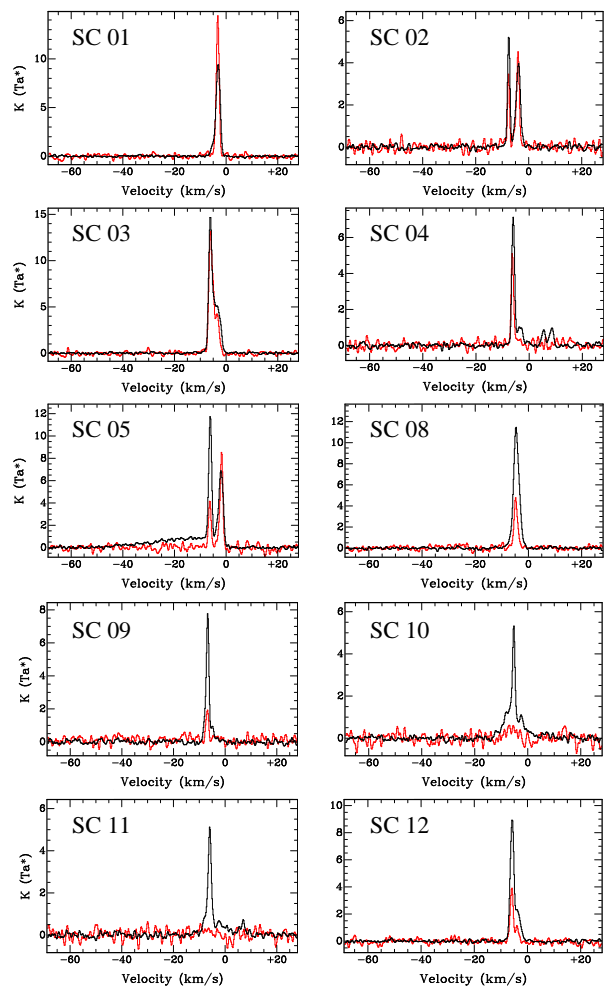}
	    \caption{
	    \twCO ($J$=1--0; black) and \thCO ($J$=1--0;
	    red, multiplied by a factor of four) spectra of SCs 01--12.
	    We do not show the spectra of SCs 06 and 07 because the
	    spectra of the two SCs are similar to those of the shocked
	    clumps D and E (see Figures~2 and 3).
	    \label{fig5}}
\end{figure}

\section{RESULTS}
\subsection{Molecular Emission}
\subsubsection{Global MC Distribution and the Shocked Molecular Clumps}

\twCO~($J$=1--0), compared to \thCO~($J$=1--0), usually show broad wings
or asymmetric profiles that result from the shock perturbation of the
molecular gas. Combining the emission of
\twCO~($J$=1--0) and \thCO~($J$=1--0), we investigate the overall
nature of the ambient molecular gas toward SNR IC~443.

Most of the \twCO\ and \thCO\ emission in the vicinity of SNR IC~443 arises
from the velocity intervals of $-50$~km~s$^{-1}$ to 10~km~s$^{-1}$ and
$-10$~km~s$^{-1}$ to 10~km~s$^{-1}$, respectively.
In Figure~1, we find that the molecular gas in the interval
of $-10$~km~s$^{-1}$ to 10~km~s$^{-1}$ has a complicated
distribution. On the contrary, the shocked gas in the interval
of $-50$~km~s$^{-1}$ to $-9$~km~s$^{-1}$ and 0~km~s$^{-1}$ to 4~km~s$^{-1}$,
indicated by the broadened \twCO,
is mainly in the southeast and the center of the remnant,
respectively. C$^{18}$O~($J$=1--0) emission is generally too weak
to examine its spatial distribution, except that only some points with
strong \thCO\ emission were detected in the faint C$^{18}$O line, e.g.,
shocked clump G and small clouds (SCs; named by Lee et al. 2012)
01, 03, 04, and 05.

\begin{figure*}
            \includegraphics[trim=0mm 0mm 0mm 140mm,scale=0.31,angle=0]{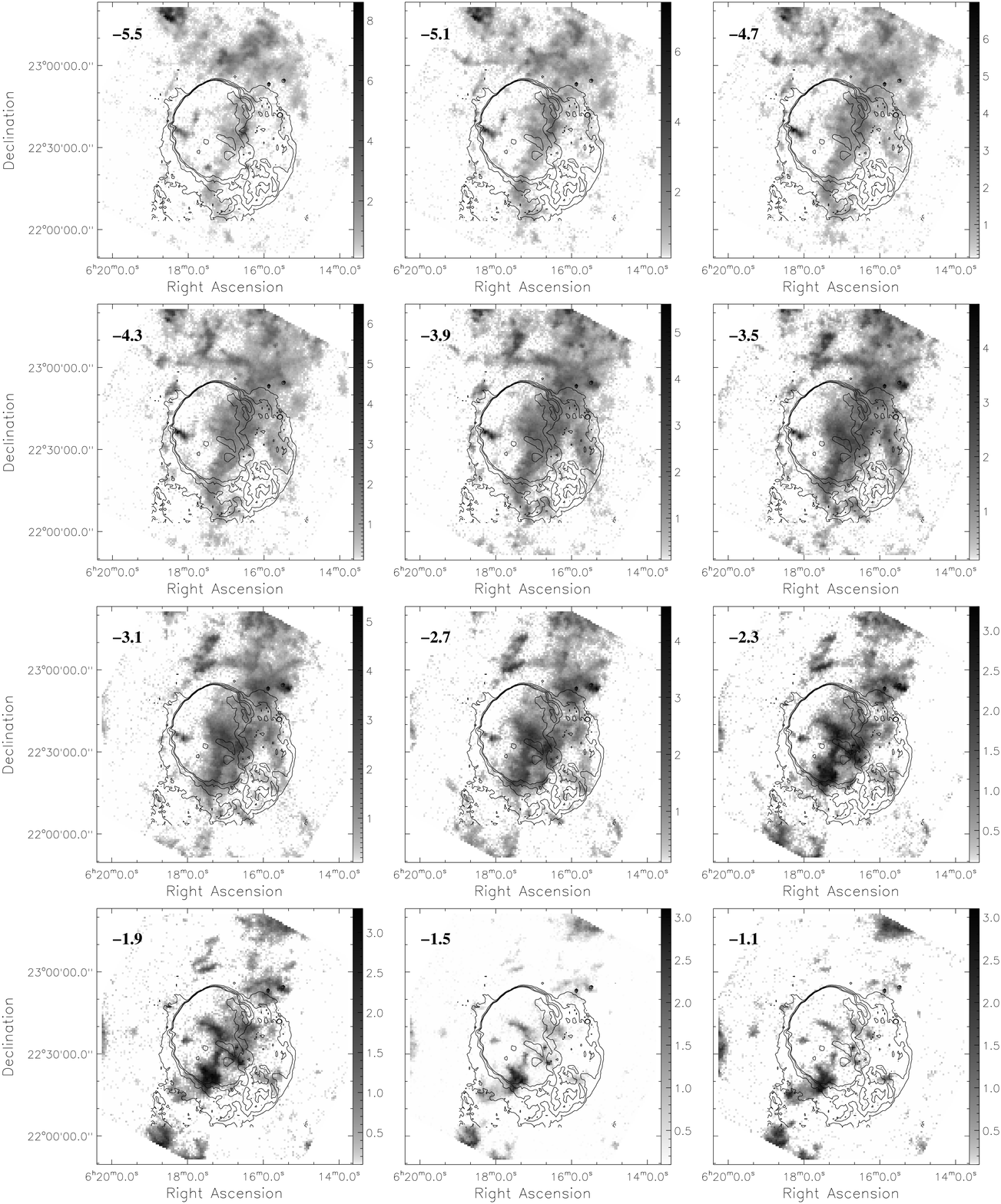}
	    \caption{
	    \twCO ($J$=1--0) intensity channel maps in the velocity range between
	    $-5.5$~km~s$^{-1}$ and $-1.1$~km~s$^{-1}$, overlaid with the 1.4~GHz
	    radio continuum emission contours. Note that the intensity scales
	    are different from each other.
	    \label{fig6}}
\end{figure*}

We present the \twCO\ and \thCO\ spectra of shocked clumps B--G
in Figure~2. The \thCO\ lines show narrow Gaussian profiles
(the red lines in Figure~2), while the \twCO\ lines show broad wing profiles
that represent the gas perturbed by the remnant's shock
(the black lines in Figure~2).
In Table~1, we list the \thCO\ parameters of shocked
clumps B--G. We find that the LSR velocities of the
\thCO\ emission are around $-4$~km~s$^{-1}$,
except for clumps C and D (Table~1). For shocked clump C,
we do not find the prominent \thCO\ emission
(Figure~2). On the other hand, the
\thCO\ emission of shocked clump D peaks at $-7.8$~km~s$^{-1}$, which is
consistent with the velocity of the adjacent OH 1720~MHz maser source (clump D,
$-6.85$~km~s$^{-1}$; Hewitt et al. 2006). Shocked clumps C and D seem to form
a long filament (about 13$'$) along the southeastern bright radio
border of SNR IC~443 (right panel of Figure~1).
The emission of the
shocked gas is more intense in the blueshifted side and mainly
located along the southern boundary of the remnant. In contrast, the
redshifted component is located in the center of the remnant.
For shocked clump G, the position
coincidence between the \thCO\ line ($-4.4$~km~s$^{-1}$; Table~1 and Figure~2)
and OH maser emission ($-4.5$~km~s$^{-1}$; Hewitt et al. 2006) toward it
indicates that the gas emission with a similar LSR velocity arises from the
same physical environment.

To investigate the detailed structures of the MCs toward
SNR IC~443, we made an intensity weighted velocity (the
first moment) map of \thCO\ emission in the velocity range
of $-10$~km~s$^{-1}$ to 10~km~s$^{-1}$ (Figure~3, left panel).
We identify five \thCO\ components in the FOV:
$-7$~km~s$^{-1}$ ($\VLSR$=$-10$~km~s$^{-1}$ to $-5$~km~s$^{-1}$, blue),
$-4$~km~s$^{-1}$ ($\VLSR$=$-6$~km~s$^{-1}$ to $-1$~km~s$^{-1}$, cyan),
3~km~s$^{-1}$ ($\VLSR$=1.5~km~s$^{-1}$ to 5~km~s$^{-1}$, green),
5~km~s$^{-1}$ ($\VLSR$=4~km~s$^{-1}$ to 7~km~s$^{-1}$, yellow),
and 8~km~s$^{-1}$ ($\VLSR$=7~km~s$^{-1}$ to 9~km~s$^{-1}$, orange).
The typical spectra for different MC components are shown
in Figure~4. The extraction region for these spectra are denoted
in the left panel of Figure~3. The spatial distribution of the five
velocity components
can also be discerned in the first-moment maps of \thCO\ (Figure~3).
The $-7$~km~s$^{-1}$ \thCO\ gas is mainly located to the
north of the remnant. The 5~km~s$^{-1}$ gas can be divided
into two parts: the eastern part is located in the southwest
of the remnant and the western part along the western radio
halo of the remnant.
The 3~km~s$^{-1}$ MCs located to the northeast of the remnant
and the 8~km~s$^{-1}$ MCs located to the east of the remnant seem
not to be associated with the SNR.

\begin{figure*}
\centerline{\includegraphics[trim=0mm 0mm 0mm 140mm,scale=0.3,angle=0]{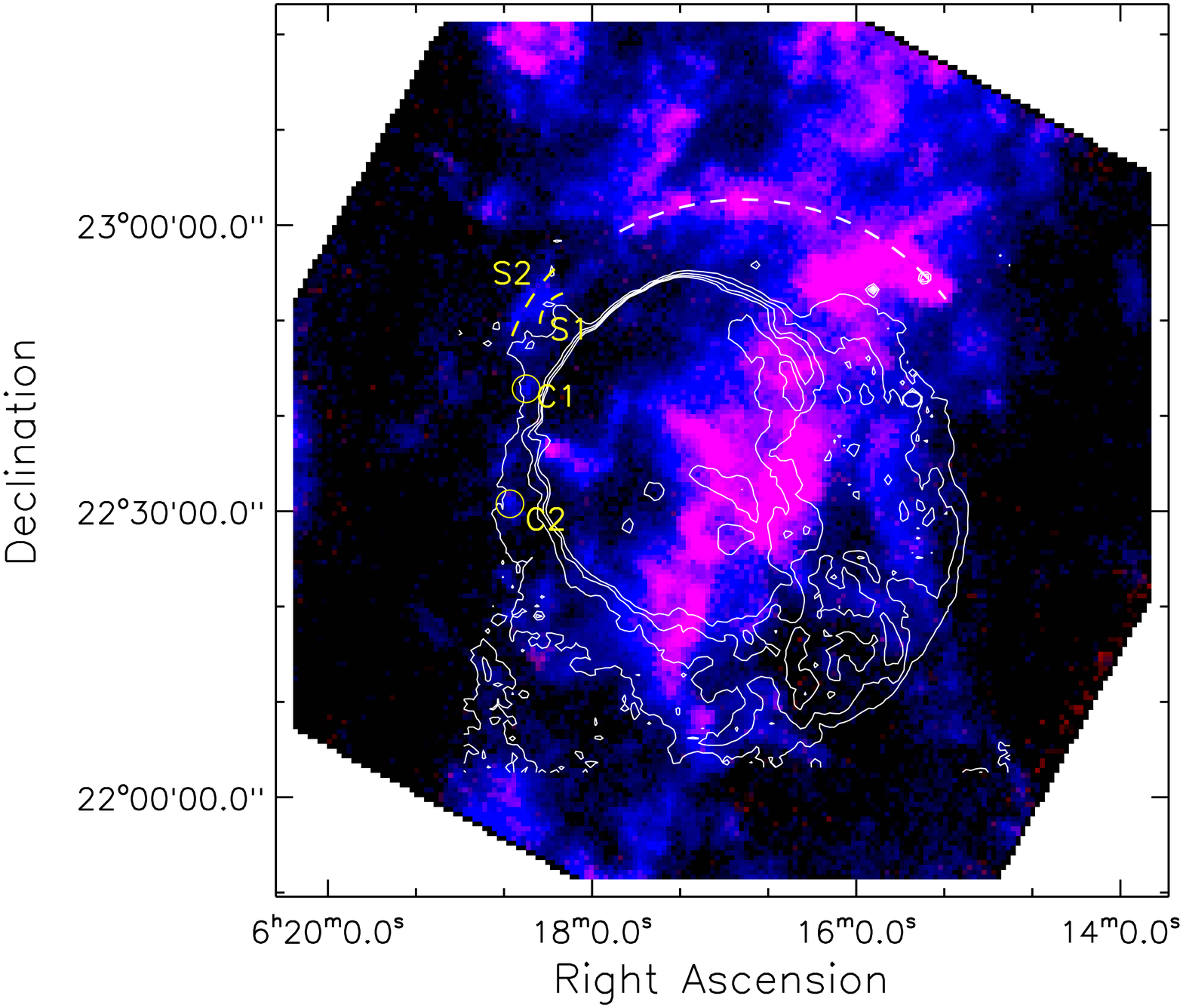}
            \includegraphics[trim=0mm 0mm 0mm 140mm,scale=0.3,angle=0]{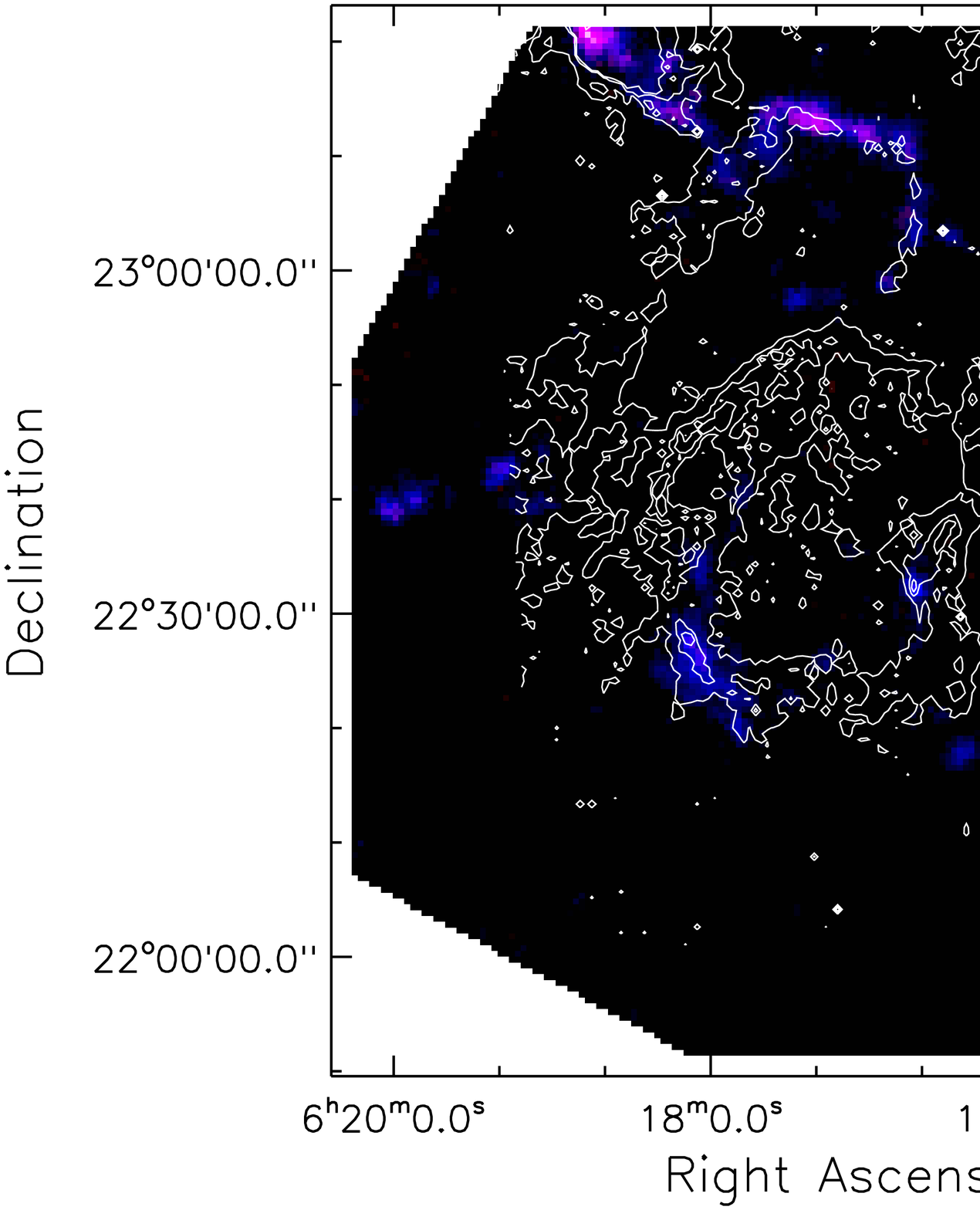}}
	    \caption{
	    Left: the \twCO\ $J$=1--0 (blue) and \thCO\ $J$=1--0 (red)
	    intensity maps in the $-6$~km~s$^{-1}$ to $-1$~km~s$^{-1}$ interval with
	    the square root scale, overlaid with the 1.4~GHz radio continuum emission
	    contours. The color scheme is
	    adjusted to highlight the faint CO structures. The northern partial shell
	    is denoted by the white
	    dashed line. The northeastern partial shell structures
	    (S1 and S2) and the eastern clumps (C1 and C2) are indicated in the image,
	    the spectra of which are shown in Figure~8. The position--velocity map along
	    the northern partial shell is shown in Figure~9.
	    Right: \twCO\ (blue) and \thCO\ (red) intensity maps in the
	    $-9.2$~km~s$^{-1}$ to $-7.8$~km~s$^{-1}$ interval with square root scale,
	    overlaid with the $Spitzer$ 24$\mu$m mid-IR emission
	    contours.
	    \label{fig7}}
\end{figure*}

\begin{figure}
	    \includegraphics[scale=1.6,angle=0]{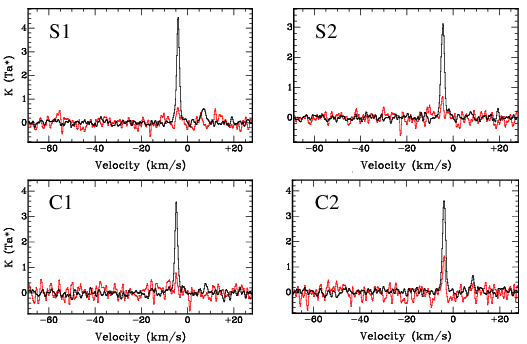}
	    \caption{
	    \twCO ($J$=1--0; black) and \thCO ($J$=1--0;
	    red, multiplied by a factor of four) spectra of S1, S2, C1, and C2
	    (see their locations in Figure~7).
	    \label{fig8}}
\end{figure}

Notably, the prominent feature in the first moment map of the \thCO\
emission is that the $-4$~km~s$^{-1}$ molecular gas extending along
southeast--northwest divides the remnant into two parts:
the multiwavelength bright part in northeast and the faint breakout
in southwest. The $-4$~km~s$^{-1}$ MCs are mostly at
the foreground of SNR IC~443 because of the evident absorption
signature in the optical (Fesen 1984; Welsh \& Sallmen 2003) and X-ray
(Troja et al. 2006) bands. Also, some partial shell-like structures are located
about 6$'$--10$'$ to the north of the remnant and are projected
concentrically with the northern radio border of SNR IC~443. To better
investigate the velocity distribution of the $-4$~km~s$^{-1}$
molecular gas, we show the close-up intensity weighted velocity map of
\thCO\ emission in the $-8$~km~s$^{-1}$ to $-1$~km~s$^{-1}$ interval in
the right panel of Figure~3, in which the location of the shocked
clumps B--G and SCs 01--12 are denoted. The spectra of these SCs are
shown in Figure~5.

\subsubsection{The Half Molecular Ring Structure Outside the SNR}
We made the \twCO\ ($J$=1--0) channel maps in
a velocity interval of $0.4$~km~s$^{-1}$ (Figure~6)
in order to illustrate any tiny structures of the gas in
the vicinity of the remnant for the $-4$~km~s$^{-1}$ MC component.
By comparing the \twCO\ images with the overlaid 1.4 GHz radio
continuum contours, there are some partial
shell structures outside the bright radio shell of the
remnant's northeastern part ($\sim5'$, approximately 2.2
pc at a typical distance of 1.5 kpc; e.g., Fesen 1984;
Braun \& Strom 1986).
The northeastern partial shell structures
can be discerned in $-4.7$~km~s$^{-1}$ to $-3.1$~km~s$^{-1}$ channel maps
(Figure~6). We also find that a few
\twCO\ clumps are precisely located along the eastern boundary
of the faint radio continuum halo (see also $-4.7$~km~s$^{-1}$ to
$-3.1$~km~s$^{-1}$ channel maps in Figure~6). The faint halo structure
appears outside the bright radio shell of SNR IC~443 and
most likely originates from the interaction of the remnant
with its environment (Lee et al. 2008).
The similar radio halo can also be found in other SNRs, e.g., SNR 3C~391
(Reynolds \& Moffett, 1993; Chen et al. 2004).
The northern
partial shell structure can be seen in $-3.9$~km~s$^{-1}$ to $-1.9$~km~s$^{-1}$
channel maps in Figure~6, similar to the structures found
in the \thCO\ first moment maps (the green arc in the right
panel of Figure~3).

\begin{deluxetable*}{cccccc}
\tabletypesize{\scriptsize}
\tablecaption{ \twCO\ ($J$=1--0) Spectral Parameters of the Partial Shell Structure}
\tablehead{
\colhead{\begin{tabular}{c}
Shells/Clumps \\
Label\\
\end{tabular}} &
\colhead{\begin{tabular}{c}
(R.A., Decl.) \\
(J2000)\\
\end{tabular}} &
\colhead{\begin{tabular}{c}
Extraction Area \\
(arcmin)\\
\end{tabular}} &
\colhead{\begin{tabular}{c}
$\VLSR$(Peak)\\
(km~s$^{-1}$)\\
\end{tabular}} &
\colhead{\begin{tabular}{c}
$T$(Peak)\\
(K)\\
\end{tabular}} &
\colhead{\begin{tabular}{c}
$W$\\
(K~km~s$^{-1}$)\\
\end{tabular}}
}
\startdata
S1 & $(\RA{06}{18}{23}.5,\Dec{22}{51}{39})$ & 3$\times$1.5 & -4.1 & 4.5 & 9.7 \\
S2 & $(\RA{06}{18}{30}.2,\Dec{22}{51}{55})$ & 7$\times$1.5 & -4.4 & 3.1 & 6.8\\
C1 & $(\RA{06}{18}{29}.3,\Dec{22}{42}{45})$ & 1$\times$1 & -4.8 & 3.6 & 7.3\\
C2 & $(\RA{06}{18}{37}.4,\Dec{22}{30}{43})$ & 1$\times$1 & -4.0 & 3.6 & 7.7
\enddata
\end{deluxetable*}

For subsequent analysis, we made the intensity maps of \twCO\ ($J$=1--0) 
and \thCO\ ($J$=1--0) in the $-6$~km~s$^{-1}$ to
$-1$~km~s$^{-1}$ interval (Figure~7, left panel) to explore the
distribution of the $-4$~km~s$^{-1}$ MCs around the SNR. In Figure~7,
the northern partial shell structure, the northeastern partial
shell structures (S1 and S2), and the eastern CO clumps (C1 and C2)
are clearly seen and form a half ring to surround SNR IC~443. 
The half-ring structure centered at
$(\RA{06}{16}{57}.1,\Dec{22}{37}{06})$ has a radius of 26$'$.
This position is close to the geometrical center of shell A,
$(\RA{06}{17}{13},\Dec{22}{37}{10})$, suggested by Lee et al. (2008).
The spectra of S1, S2, C1, and C2 are show in
Figure~8. Assuming that the line-of-sight depth of the partial shell
structures is $\sim1'$, we
estimate the mean molecular density of the partial shell structures
as $n({\rm H}_2)\sim900-1300\,{\rm cm}^{-3}$, which is consistent
with the average molecular hydrogen densities of the SCs (Lee et al.
2012). In the above calculation, we adopt the mean CO-to-H$_2$ mass conversion
factor of $1.8\E{20}$~cm$^{-2}$K$^{-1}$km$^{-1}$s (Dame et al.\ 2001)
and the distance to the SNR of 1.5~kpc. 
For clump C1 and the partial shell structures S1 and S2, we did not
use \thCO\ information to estimate the column density because of their 
weak \thCO\ emission (see Figure~8). For clump C2, assuming local 
thermodynamical equilibrium for the gas and an optically thick condition
for the \twCO\ line, we estimated the excitation temperature of the gas to be 
about 7~K and the optical depth of \thCO\ to be about 0.11. Using the relation
$N($H$_2)$=$7\times10^5N$(\thCO) (Frerking et al. 1982),
the column density of clump C2 we obtained is $\approx5\times10^{20}$~cm$^{-2}$.
Adopting the line-of-sight size of C2 $\sim1'$, the mean molecular density 
is $n({\rm H}_2)\sim400\,{\rm cm}^{-3}$, 
which is less than the mean density from the X-factor method above.
This is probably because of the low filling factor of weak \thCO\ emission 
for those clumps.
We did not calculate the mean density of the northern partial shell structure
because the structure is mixed with the foreground/background CO
emission. 

\begin{figure}
	    \includegraphics[scale=0.4,angle=0]{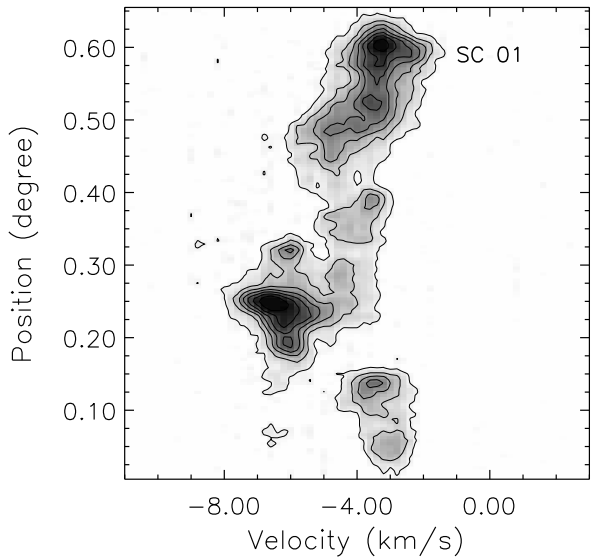}
	    \caption{
	    Position--velocity diagram of \thCO\ emission along the northern partial shell
	    structure. The position is measured from the east to the west along the
	    white dashed line (see Figure~7) with a width of 2\farcm5.
	    \label{fig9}}
\end{figure}

In Table~2, we list the parameters of the \twCO\ emission of the 
structures S1, S2, C1, and C2. The systemic LSR velocities of these structures
are between $-5$~km~s$^{-1}$ and $-4$~km~s$^{-1}$. We also made the 
position--velocity (PV) diagram of \thCO\ emission along the northern partial 
shell structure (Figure~9). We find that the velocity 
range of the shell spans from $-5$~km~s$^{-1}$ to $-2$~km~s$^{-1}$, which 
is consistent with that of the first moment maps of \thCO\ (Figure~3, 
right panel, the green arc to the north of the map). 
The additional
prominent velocity component along the line of sight in this direction is
$-8$~km~s$^{-1}$ to $-5$~km~s$^{-1}$. This can also be seen in the first
moment maps of the \thCO\ emission (Figure~3, right panel, the blue part
across the green arc to the north of the map).
We emphasize here that the velocities of the 
partial shell structures are consistent with the velocity range of the 
$-4$~km~s$^{-1}$ MC component. They may belong to the 
$-4$~km~s$^{-1}$ MCs.

\begin{figure}
	    \includegraphics[scale=0.8,angle=0]{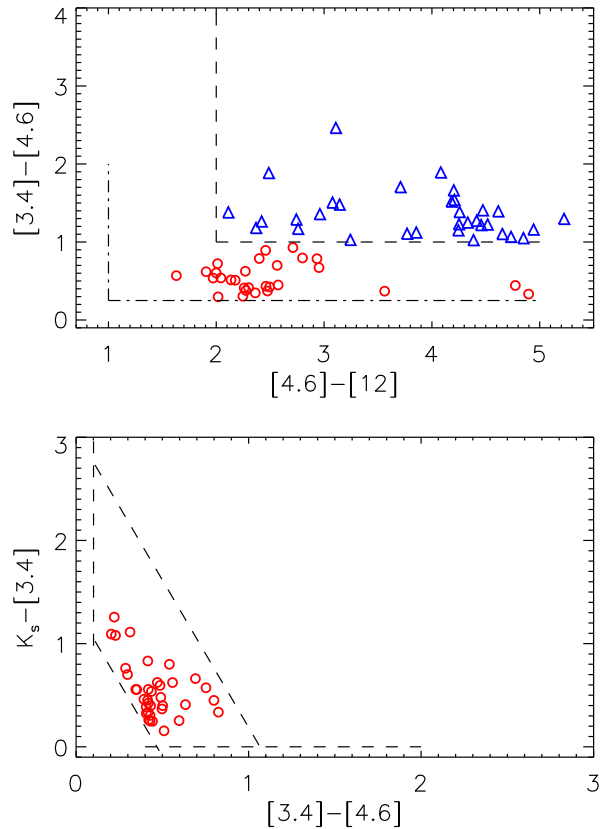}
	    \caption{
	    Upper: $WISE$ band 1, 2, and 3 color--color diagram showing the distribution
	    of YSO candidates (Class II: 29 red circles, see ID 1--29 in
	    Table~3; Class I: 33 blue triangles, see ID 30--62 in Table~3) in
	    the field of view of SNR IC~443.
	    Lower: 2MASS $K_{\rm s}$ and $WISE$ band 1 and 2 color--color diagram showing
	    the distribution of YSO candidates (Class II: 36 red circles, see ID
	    63--98 in Table~3) in the field of view of SNR IC~443.
	    \label{fig10}}
\end{figure}

\begin{figure*}
\centerline{\includegraphics[trim=0mm 0mm 0mm 140mm,scale=0.47,angle=0]{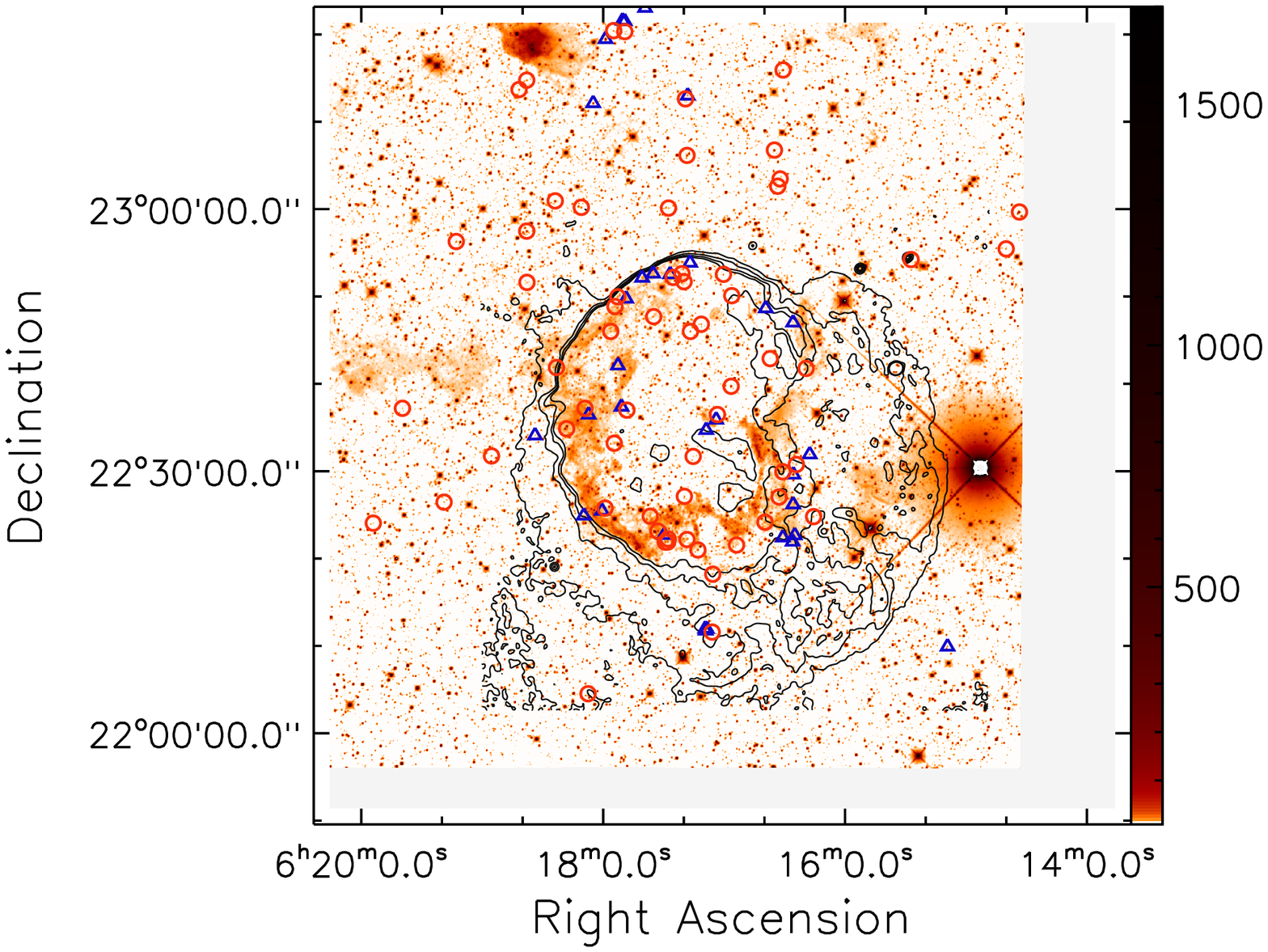}
            \includegraphics[trim=0mm 0mm 0mm 140mm,scale=0.47,angle=0]{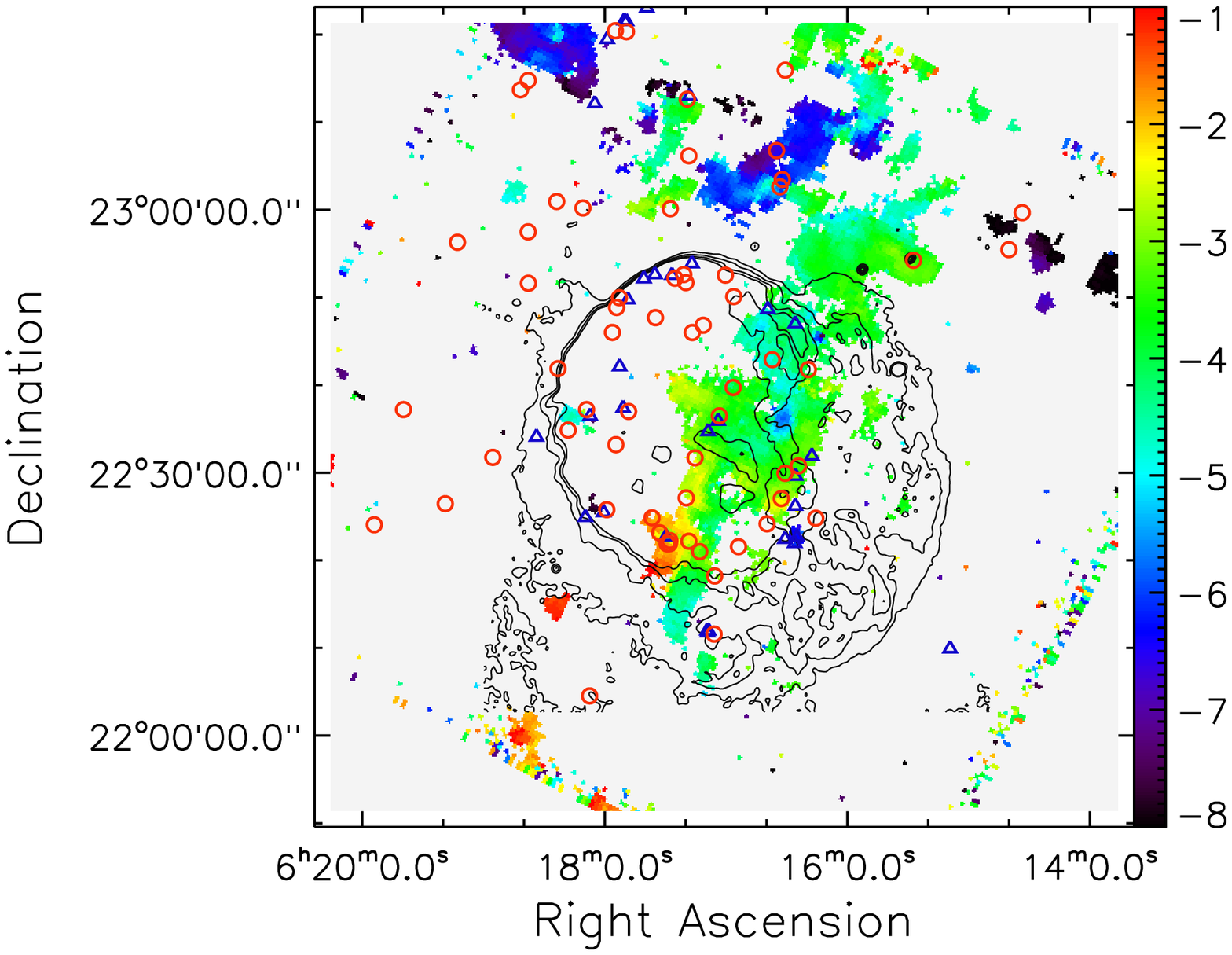}}
	    \caption{ Left: the locations of the YSO candidates superimposed on the $WISE$
	    3.4~$\mu$m emission. The black contours are from the 1.4~GHz radio continuum
	    emission. Right: the locations of the YSO candidates superimposed on the map of
	    the first moment of the \thCO\ emission in the interval of $-8$ to
	    $-1$~km~s$^{-1}$. The blue triangles are the YSO candidates of Class I,
	    and the red circles are the YSO candidates of Class II.
	    \label{fig11}}
\end{figure*}

We discovered a faint CO bubble north of the remnant and centered at
$(\RA{06}{17}{17}.3,\Dec{23}{05}{28})$ with a diameter of 16$'$,
and in the velocity range of $-9.2$~km~s$^{-1}$ to
$-7.8$~km~s$^{-1}$ (see the
right panel of Figure~7), which is coincident with the bubble bright in
the $Spitzer$ 24$\mu$m mid-IR emission.
The origin of the shell structure is unknown. In the
16$'$ cavity, we have not found any massive stars in the SIMBAD Astronomical
Database.
However, a YSO candidate (J061718.39+230609.0, Class II) is found to be
located near the center of the CO bubble.

\subsection{YSO Distribution Near SNR IC~443}

Using the IR data from the 2MASS and $WISE$ surveys, we searched
for the disk-bearing young stellar population in the field of IC~443.
The dusty circumstellar disks can emit IR emission, which causes
the IR colors of the disk-bearing young stars to be distinctly different
from those of diskless objects. However, diskless young stars cannot
be distinguished from unrelated field objects based on IR colors
alone. Therefore, we only investigate
the disk-bearing young stars in this study.

We selected candidate disk-bearing young stars using the criteria
described in Koenig et al. (2012). The inventory of the young disk
population is first based on the $WISE$ [3.4]--[4.6] versus [4.6]--[12]
color--color diagram (Figure~10, upper panel). We removed the contaminations
from the extragalactic sources
(star-forming galaxy and active galactic nuclei), the blobs of the shock
emission, and the
resolved polycyclic aromatic hydrocarbon emission objects
according to their locations in the
[3.4]--[4.6] versus [4.6]--[12] color--color diagram and their
$WISE$ photometry (see the detailed description in Koenig et al. 2012).
For the sources that were not detected in the $WISE$ [12] band,
their dereddened photometry in the $WISE$ [3.4] and [4.6] bands,
in combination of the dereddened 2MASS $K_{\rm s}$ photometry,
was used to construct the $K_{\rm s}$--[3.4] versus [3.4]--[4.6]
color--color diagram (Figure~10, lower panel). The extinction used to deredden
the photometry was
estimated from its location in the $J$--$H$ versus $H$--$K_{\rm s}$
color--color diagram as described in Fang et al. (2013).

In the $80' \times 80'$ area centered at ($l$=189\fdg0,$b$=3\fdg0),
we identified 98 (33 Class I and 65 Class II) potential YSOs
from the $WISE$+2MASS color and the magnitude cuts imposed on the 15,077 sources
with good S/N photometry. Figure~11 shows the locations of the YSO
candidates on the $WISE$ 3.4~$\mu$m map and the first moment of
the \thCO\ emission map, respectively.
Sixty-two (24 Class I and 38 Class II) YSO candidates are located
within the radio halo of SNR IC~443.
We show the color--color diagrams
of the YSO candidates in the FOV of SNR IC~443 (Figure~10) and
list the information of the YSO candidates in Table~3.
ID 1--29 are Class II YSO candidates (29 red circles in the upper panel
of Figure~10) based on the $WISE$ band 1, 2, and 3 color--color diagram.
ID 30--62 are Class I YSO candidates (33 blue triangles in the upper
panel of Figure~10) based on the $WISE$ band 1, 2, and 3 color--color
diagram. ID 63--98 are Class II
YSO candidates (36 red circles in the lower panel of Figure~10) based on 2MASS
$K_{\rm s}$ and the $WISE$ band 1 and 2 color--color diagram.
We also estimate the spectral slope, $\alpha$, of the 98 sources
from the IR spectral energy distributions (SEDs) to classify the
evolutionary stages of these sources (Greene et al. 1994). We find
that it is consistent with the result of the IR color--color method (Table~3).

The YSO candidates (Figure~11) are mostly
distributed along the bright radio boundary, but are absent in the
southwestern breakout portion of the remnant.

\section{DISCUSSION}
\subsection{Association of SNR IC~443 with the $\VLSR
 \sim-4$~km~s$^{-1}$ MCs}

The 1720~MHz OH maser is an important signpost of the interaction between 
SNRs and MCs (Lockett et al. 1999). For SNR IC~443, three OH masers were 
reported in Hewitt et al. (2006), and the locations of these three OH
masers are close to the shocked clumps SC~05, D, and G,
respectively. The three OH masers have velocities of
$-6.14$~km~s$^{-1}$, $-6.85$~km~s$^{-1}$, and $-4.55$~km~s$^{-1}$, 
respectively. These velocities represent the systemic velocities of the
shocked gas that varies in a few kilometers per second. 
We found that there are two narrow \thCO\ components with 
velocities centered at $-6.4$~km~s$^{-1}$ and $-2.1$~km~s$^{-1}$
in the shocked clump SC~05 and one narrow \thCO\ component centered at
$-7.8$~km~s$^{-1}$ in the shocked clump D. 
As two OH masers are adjacent to the two shocked clumps, 
the $-6.4$~km~s$^{-1}$ and $-7.8$~km~s$^{-1}$ components of the \thCO\ 
emission are likely due to the preshock gas while the $-2.1$~km~s$^{-1}$ 
component might be from the foreground/background MC along the line of 
sight. The $-4$~km~s$^{-1}$ MCs seems to divide the remnant into the
multiwavelength bright part in northeast and the faint 
breakout part in southwest. It is noteworthy that the shocked clump G 
is projected 
around the centroid of the $-4$~km~s$^{-1}$ MCs extending along the 
southeast--northwest. Figure~12 shows PV diagrams of \twCO\ 
and \thCO\ across the shock clump G along the southwest--northeast 
(the short arrow in the right panel of Figure~3).
The LSR velocity coincidence between the \thCO\ line and OH maser
emission toward the shocked clump G (see Section 3.1.1) indicates
that the $-4$~km~s$^{-1}$ molecular gas could be associated with
SNR IC~443 (Figure~12). 

\begin{longtable*}{ccccccccccc}
\caption{The Infrared Photometric Magnitudes of the YSO Candidates toward SNR IC~443}\\
\hline\hline
 ID  &  Catalog Number & $J$(1.25$\mu$m) & $H$(1.65$\mu$m) & $K_{\rm S}$(2.17$\mu$m) & W1(3.4$\mu$m) & W2(4.6$\mu$m) & W3(12$\mu$m) & W4(22$\mu$m) & Av & $\alpha$\\
  & & (mag) & (mag) & (mag) & (mag)  &  (mag) &  (mag) &  (mag)  & (mag) &  \\
\hline
\endfirsthead
\caption{continued.}\\
\hline\hline
 ID & Catalog Number & $J$(1.25$\mu$m) & $H$(1.65$\mu$m) & $K_{\rm S}$(2.17$\mu$m) & W1(3.4$\mu$m) & W2(4.6$\mu$m) & W3(12$\mu$m) & W4(22$\mu$m) & Av & $\alpha$ \\
  & & (mag) & (mag) & (mag) & (mag)  &  (mag) &  (mag) &  (mag) & (mag) &  \\
\hline
\endhead
\hline
\endfoot
  1 &  J061635.03+230645.3 & 14.31$\pm$0.03  &   13.25$\pm$0.04 &   12.43$\pm$0.02 &   11.57$\pm$0.02  &  11.00$\pm$0.02  &   9.37$\pm$0.05  &  7.10$\pm$0.10  &      0.86      &   -1.22   \\ [0.3pt]   
  2 &  J061837.94+231445.7 & 15.32$\pm$0.06  &   14.14$\pm$0.06 &   13.61$\pm$0.04 &   12.84$\pm$0.03  &  12.12$\pm$0.03  &  10.11$\pm$0.08  &  8.81$\pm$0.00  &      4.97      &   -1.06   \\ [0.3pt]   
  3 &  J061841.76+231341.3 & 14.60$\pm$0.04  &   13.58$\pm$0.03 &   12.91$\pm$0.03 &   12.08$\pm$0.03  &  11.57$\pm$0.02  &   9.39$\pm$0.06  &  7.25$\pm$0.14  &      1.48      &   -0.97   \\ [0.3pt]   
  4 &  J061630.72+231555.9 & 14.47$\pm$0.04  &   13.36$\pm$0.04 &   12.95$\pm$0.03 &   12.21$\pm$0.03  &  11.59$\pm$0.02  &   9.68$\pm$0.08  &  7.71$\pm$0.18  &      3.93      &   -1.17   \\ [0.3pt]   
  5 &  J061754.79+232025.6 & 16.34$\pm$0.11  &   14.98$\pm$0.09 &   14.11$\pm$0.04 &   13.15$\pm$0.03  &  12.22$\pm$0.03  &   9.51$\pm$0.07  &  6.86$\pm$0.08  &      4.68      &   -0.43   \\ [0.3pt]   
  6 &  J061741.61+232436.4 & 15.95$\pm$0.07  &   14.55$\pm$0.06 &   13.60$\pm$0.03 &   12.67$\pm$0.03  &  11.88$\pm$0.03  &   9.48$\pm$0.05  &  7.30$\pm$0.13  &      4.56      &   -0.71   \\ [0.3pt]   
  7 &  J061719.17+231237.1 & 14.98$\pm$0.04  &   13.73$\pm$0.04 &   12.92$\pm$0.03 &   11.82$\pm$0.02  &  11.28$\pm$0.02  &   9.31$\pm$0.07  &  6.69$\pm$0.07  &      3.52      &   -1.01   \\ [0.3pt]   
  8 &  J061718.39+230609.0 & 13.81$\pm$0.03  &   13.08$\pm$0.03 &   12.35$\pm$0.02 &   11.31$\pm$0.02  &  10.68$\pm$0.02  &   8.41$\pm$0.03  &  6.37$\pm$0.07  &      0.00      &   -0.72   \\ [0.3pt]   
  9 &  J061632.20+230327.2 & 15.49$\pm$0.05  &   14.23$\pm$0.05 &   13.45$\pm$0.03 &   12.61$\pm$0.03  &  11.82$\pm$0.02  &   8.89$\pm$0.04  &  6.54$\pm$0.06  &      4.21      &   -0.43   \\ [0.3pt]   
 10 &  J061705.93+221134.7 & 14.23$\pm$0.03  &   13.60$\pm$0.04 &   13.40$\pm$0.04 &   13.07$\pm$0.03  &  12.66$\pm$0.04  &  10.40$\pm$0.10  &  8.39$\pm$0.00  &      0.00      &   -1.17   \\ [0.3pt]   
 11 &  J061837.93+225728.6 & 13.95$\pm$0.03  &   13.07$\pm$0.03 &   12.48$\pm$0.02 &   11.69$\pm$0.02  &  11.27$\pm$0.02  &   8.77$\pm$0.03  &  6.80$\pm$0.10  &      0.15      &   -0.82   \\ [0.3pt]   
 12 &  J061440.06+225527.0 & 14.93$\pm$0.04  &   13.87$\pm$0.04 &   13.31$\pm$0.03 &   12.83$\pm$0.03  &  12.48$\pm$0.04  &  10.12$\pm$0.07  &  7.35$\pm$0.14  &      4.20      &   -1.19   \\ [0.3pt]   
 13 &  J061656.51+223943.2 & 14.01$\pm$0.04  &   13.19$\pm$0.04 &   12.50$\pm$0.03 &   11.18$\pm$0.02  &  10.66$\pm$0.02  &   8.53$\pm$0.03  &  5.96$\pm$0.06  &      0.00      &   -0.74   \\ [0.3pt]   
 14 &  J061716.73+224559.9 & 13.91$\pm$0.03  &   13.32$\pm$0.03 &   12.86$\pm$0.02 &   12.23$\pm$0.03  &  11.86$\pm$0.03  &   8.30$\pm$0.03  &  5.58$\pm$0.04  &      0.00      &   -0.30   \\ [0.3pt]   
 15 &  J061700.34+225232.0 & 14.65$\pm$0.04  &   13.77$\pm$0.04 &   13.35$\pm$0.03 &   12.97$\pm$0.03  &  12.60$\pm$0.03  &  10.12$\pm$0.07  &  6.39$\pm$0.06  &      2.55      &   -1.11   \\ [0.3pt]   
 16 &  J061653.86+222132.7 & 15.23$\pm$0.05  &   14.52$\pm$0.06 &   14.24$\pm$0.05 &   13.90$\pm$0.04  &  13.23$\pm$0.06  &  10.28$\pm$0.10  &  8.34$\pm$0.33  &      0.73      &   -0.62   \\ [0.3pt]   
 17 &  J061630.86+222955.8 & 13.73$\pm$0.03  &   13.05$\pm$0.04 &   12.91$\pm$0.03 &   12.68$\pm$0.03  &  12.30$\pm$0.03  &  10.03$\pm$0.10  &  8.70$\pm$0.00  &      0.75      &   -1.24   \\ [0.3pt]   
 18 &  J061748.23+223659.3 & 13.98$\pm$0.03  &   13.65$\pm$0.05 &   13.47$\pm$0.04 &   13.06$\pm$0.03  &  12.61$\pm$0.03  &  10.04$\pm$0.08  &  7.26$\pm$0.11  &      0.00      &   -0.92   \\ [0.3pt]   
 19 &  J061619.31+224145.5 & 15.92$\pm$0.07  &   14.95$\pm$0.07 &   14.77$\pm$0.08 &   14.04$\pm$0.05  &  13.71$\pm$0.06  &   8.81$\pm$0.05  &  7.09$\pm$0.12  &      3.45      &    0.41   \\ [0.3pt]   
 20 &  J061759.05+222546.6 & 12.97$\pm$0.03  &   12.74$\pm$0.03 &   12.60$\pm$0.03 &   12.37$\pm$0.03  &  12.07$\pm$0.04  &  10.05$\pm$0.09  &  7.78$\pm$0.21  &      0.00      &   -1.42   \\ [0.3pt]   
 21 &  J061837.82+225136.7 & \nodata   		   &    \nodata			  &	 \nodata 			 	 &   14.30$\pm$0.04  &  13.86$\pm$0.05  &   9.08$\pm$0.05  &  7.19$\pm$0.14  &       \nodata  &    1.17   \\ [0.3pt]   
 22 &  J061823.79+230056.5 & 13.25$\pm$0.04  &   12.65$\pm$0.05 &   12.05$\pm$0.03 &   11.28$\pm$0.02  &  10.86$\pm$0.02  &   8.56$\pm$0.03  &  6.34$\pm$0.06  &       \nodata  &   -0.94   \\ [0.3pt]   
 23 &  J061727.79+222214.7 & 15.02$\pm$0.04  &   13.84$\pm$0.03 &   13.26$\pm$0.03 &   12.63$\pm$0.03  &  11.83$\pm$0.03  &   9.03$\pm$0.11  &  6.68$\pm$0.09  &      5.32      &   -0.63   \\ [0.3pt]   
 24 &  J061855.42+223144.8 & 15.70$\pm$0.06  &   14.62$\pm$0.07 &   13.73$\pm$0.04 &   12.77$\pm$0.03  &  12.17$\pm$0.03  &  10.17$\pm$0.10  &  8.16$\pm$0.33  &      0.56      &   -0.94   \\ [0.3pt]   
 25 &  J061912.92+225617.5 & 14.94$\pm$0.05  &   14.10$\pm$0.05 &   13.69$\pm$0.04 &   12.94$\pm$0.03  &  12.40$\pm$0.03  &  10.36$\pm$0.10  &  7.90$\pm$0.21  &      2.27      &   -1.09   \\ [0.3pt]   
 26 &  J061719.72+222707.3 & 14.69$\pm$0.04  &   13.72$\pm$0.03 &   13.28$\pm$0.03 &   12.50$\pm$0.03  &  11.80$\pm$0.02  &   9.23$\pm$0.04  &  7.90$\pm$0.28  &      3.20      &   -0.70   \\ [0.3pt]   
 27 &  J061632.71+222702.3 & 13.54$\pm$0.03  &   13.15$\pm$0.04 &   13.03$\pm$0.02 &   12.80$\pm$0.03  &  12.36$\pm$0.03  &   9.90$\pm$0.10  &  7.77$\pm$0.22  &      0.00      &   -1.07   \\ [0.3pt]   
 28 &  J061713.00+222058.8 & 14.24$\pm$0.03  &   13.34$\pm$0.03 &   12.94$\pm$0.02 &   12.61$\pm$0.03  &  12.31$\pm$0.03  &  10.06$\pm$0.09  &  8.45$\pm$0.00  &      2.60      &   -1.31   \\ [0.3pt]   
 29 &  J061734.95+224741.7 & 15.12$\pm$0.05  &   14.40$\pm$0.06 &   14.22$\pm$0.05 &   13.71$\pm$0.04  &  12.81$\pm$0.07  &  10.36$\pm$0.10  &  7.56$\pm$0.22  &      0.21      &   -0.68   \\ [0.3pt]   
 30 &  J061805.00+231208.0 & 15.66$\pm$0.07  &   14.47$\pm$0.07 &   13.52$\pm$0.04 &   11.93$\pm$0.02  &  10.90$\pm$0.02  &   7.65$\pm$0.02  &  4.78$\pm$0.03  &      1.64      &    0.29   \\ [0.3pt]   
 31 &  J061749.29+232130.7 & 14.79$\pm$0.04  &   14.27$\pm$0.05 &   13.94$\pm$0.04 &   12.83$\pm$0.03  &  11.73$\pm$0.02  &   7.96$\pm$0.02  &  5.78$\pm$0.04  &      0.00      &    0.47   \\ [0.3pt]   
 32 &  J061758.83+231928.2 & 15.20$\pm$0.04  &   13.31$\pm$0.04 &   11.99$\pm$0.02 &   10.81$\pm$0.03  &   9.63$\pm$0.02  &   7.26$\pm$0.02  &  4.79$\pm$0.03  &      8.60      &   -0.49   \\ [0.3pt]   
 33 &  J061739.23+232302.4 & 16.53           &   15.51          &   14.19$\pm$0.07 &   11.88$\pm$0.03  &  10.00$\pm$0.02  &   7.51$\pm$0.02  &  4.83$\pm$0.04  &       \nodata  &    0.72   \\ [0.3pt]   
 34 &  J061718.10+231300.1 &   \nodata       &    \nodata       &   \nodata        &   13.71$\pm$0.03  &  12.24$\pm$0.03  &   9.09$\pm$0.04  &  6.10$\pm$0.05  &       \nodata  &    0.46   \\ [0.3pt]   
 35 &  J061750.41+232135.6 & 16.76$\pm$0.15  &   15.52$\pm$0.16 &   14.83$\pm$0.09 &   13.53$\pm$0.03  &  12.24$\pm$0.03  &   9.49$\pm$0.06  &  6.42$\pm$0.06  &      5.73      &   -0.08   \\ [0.3pt]   
 36 &  J061742.01+232423.4 & 17.36           &   15.44          &   15.19$\pm$0.13 &   13.87$\pm$0.03  &  12.61$\pm$0.03  &  10.18$\pm$0.10  &  8.23$\pm$0.41  &       \nodata  &   -0.12   \\ [0.3pt]   
 37 &  J061709.15+221145.0 &   \nodata       &     \nodata      &     \nodata      &   15.46$\pm$0.09  &  13.92$\pm$0.06  &   9.71$\pm$0.06  &  8.25$\pm$0.31  &       \nodata  &    1.35   \\ [0.3pt]   
 38 &  J061729.73+222242.5 &   \nodata       &      \nodata     &      \nodata     &   13.72$\pm$0.04  &  11.26$\pm$0.03  &   8.15$\pm$0.03  &  5.17$\pm$0.04  &       \nodata  &    1.00   \\ [0.3pt]   
 39 &  J061625.78+224701.3 &    \nodata      &       \nodata    &     \nodata      &   15.63$\pm$0.10  &  14.53$\pm$0.11  &   9.87$\pm$0.08  &  7.70$\pm$0.21  &       \nodata  &    1.46   \\ [0.3pt]   
 40 &  J061708.88+223443.3 &    \nodata      &       \nodata    &     \nodata      &   15.40$\pm$0.06  &  14.13$\pm$0.07  &   9.72$\pm$0.06  &  8.65$\pm$0.00  &       \nodata  &    1.36   \\ [0.3pt]   
 41 &  J061509.13+220956.3 & 16.52$\pm$0.12  &   15.41$\pm$0.10 &   14.76$\pm$0.09 &   12.91$\pm$0.03  &  11.55$\pm$0.02  &   8.59$\pm$0.03  &  6.36$\pm$0.07  &      3.03      &    0.40   \\ [0.3pt]   
 42 &  J061809.61+222455.5 &   \nodata       &        \nodata   &     \nodata      &   15.34$\pm$0.08  &  14.09$\pm$0.08  &   9.75$\pm$0.09  &  8.35$\pm$0.00  &       \nodata  &    1.28   \\ [0.3pt]   
 43 &  J061626.00+222156.5 &    \nodata      &      \nodata     &     \nodata      &   15.09$\pm$0.06  &  13.87$\pm$0.07  &   9.35$\pm$0.06  &  7.55$\pm$0.20  &       \nodata  &    1.42   \\ [0.3pt]   
 44 &  J061735.17+225237.3 &   \nodata       &      \nodata     &     \nodata      &   14.79$\pm$0.07  &  13.76$\pm$0.09  &   9.38$\pm$0.11  &  7.70$\pm$0.41  &       \nodata  &    1.20   \\ [0.3pt]   
 45 &  J061624.95+222239.6 &    \nodata      &     \nodata      &    \nodata       &   15.86$\pm$0.10  &  14.56$\pm$0.09  &   9.33$\pm$0.07  &  7.75$\pm$0.24  &       \nodata  &    2.03   \\ [0.3pt]   
 46 &  J061807.22+223628.7 &   \nodata       &     \nodata      &     \nodata      &   14.59$\pm$0.11  &  13.07$\pm$0.07  &   8.88$\pm$0.05  &  7.82$\pm$0.30  &       \nodata  &    1.32   \\ [0.3pt]   
 47 &  J061709.33+221159.0 &     \nodata     &       \nodata    &     \nodata      &   15.62$\pm$0.09  &  14.56$\pm$0.10  &   9.82$\pm$0.07  &  8.51$\pm$0.36  &       \nodata  &    1.50   \\ [0.3pt]   
 48 &  J061625.75+222612.2 &    \nodata      &      \nodata     &      \nodata     &   15.60$\pm$0.10  &  14.19$\pm$0.09  &   9.72$\pm$0.09  &  8.01$\pm$0.28  &       \nodata  &    1.49   \\ [0.3pt]   
 49 &  J061625.62+222937.0 & 16.80$\pm$0.15  &   14.32$\pm$0.06 &   12.74$\pm$0.02 &   11.05$\pm$0.02  &   9.55$\pm$0.02  &   6.47$\pm$0.02  &  4.16$\pm$0.02  &     14.73      &    0.18   \\ [0.3pt]   
 50 &  J061617.53+223155.7 & 16.87           &   15.36$\pm$0.11 &   13.64$\pm$0.04 &   12.33$\pm$0.02  &  10.95$\pm$0.02  &   8.84$\pm$0.03  &  5.97$\pm$0.05  &       \nodata  &   -0.24   \\ [0.3pt]   
 51 &  J061726.78+225232.8 &  \nodata        &    \nodata       &     \nodata      &   15.09$\pm$0.09  &  13.95$\pm$0.09  &   9.70$\pm$0.08  &  7.41$\pm$0.22  &       \nodata  &    1.16   \\ [0.3pt]   
 52 &  J061748.60+224945.5 &   \nodata       &      \nodata     &          \nodata &   15.50$\pm$0.10  &  13.84$\pm$0.09  &   9.64$\pm$0.07  &  7.53$\pm$0.23  &       \nodata  &    1.42   \\ [0.3pt]   
 53 &  J061750.93+223723.3 &    \nodata      &     \nodata      &          \nodata &   15.26$\pm$0.08  &  13.88$\pm$0.07  &   9.62$\pm$0.06  &  7.49$\pm$0.16  &       \nodata  &    1.30   \\ [0.3pt]   
 54 &  J061740.69+225208.9 & 16.26           &   15.78          &   15.39$\pm$0.16 &   13.81$\pm$0.05  &  12.69$\pm$0.04  &   8.84$\pm$0.05  &  6.25$\pm$0.09  &       \nodata  &    0.74   \\ [0.3pt]   
 55 &  J061639.32+224839.1 &   \nodata       &      \nodata     &       \nodata    &   15.48$\pm$0.07  &  14.08$\pm$0.06  &   9.47$\pm$0.04  &  6.74$\pm$0.08  &       \nodata  &    1.59   \\ [0.3pt]   
 56 &  J061800.63+222529.0 &   \nodata       &       \nodata    &      \nodata     &   15.35$\pm$0.09  &  13.46$\pm$0.06  &   9.38$\pm$0.06  &  8.14$\pm$0.53  &       \nodata  &    1.45   \\ [0.3pt]   
 57 &  J061716.87+225352.2 &   \nodata 			 &   \nodata 				&    	   \nodata	 &   15.73$\pm$0.09  &  14.57$\pm$0.10  &   9.63$\pm$0.06  &  6.49$\pm$0.08  &       \nodata  &    1.72   \\ [0.3pt]   
 58 &  J061630.87+222225.3 &  \nodata				 &     \nodata 			& 	    \nodata 	 &   15.58$\pm$0.08  &  14.37$\pm$0.08  &   9.91$\pm$0.08  &  8.59$\pm$0.43  &       \nodata  &    1.36   \\ [0.3pt]   
 59 &  J061833.83+223406.1 &   \nodata 			 &     \nodata 			& 	    \nodata		 &   13.74$\pm$0.03  &  12.57$\pm$0.03  &   9.81$\pm$0.07  &  7.51$\pm$0.17  &       \nodata  &   -0.02   \\ [0.3pt]   
 60 &  J061752.68+224206.8 & 17.14$\pm$0.22  &   15.93$\pm$0.17 &   15.76$\pm$0.21 &   14.74$\pm$0.09  &  13.52$\pm$0.09  &   9.27$\pm$0.04  &  8.33$\pm$0.00  &      5.78      &    0.64   \\ [0.3pt]   
 61 &  J061708.62+221156.3 &   	 \nodata 		 &   	   \nodata 		&    	     \nodata &   15.56$\pm$0.09  &  14.51$\pm$0.09  &   9.66$\pm$0.06  &  7.39$\pm$0.12  &       \nodata  &    1.58   \\ [0.3pt]   
 62 &  J061703.93+223554.7 &     \nodata 		 &   	   \nodata 		&    	   \nodata	 &   14.97$\pm$0.08  &  13.27$\pm$0.06  &   9.56$\pm$0.07  &  8.32$\pm$0.30  &       \nodata  &    1.05   \\ [0.3pt]   
 63 &  J061433.46+225940.2 & 11.75$\pm$0.03  &   11.22$\pm$0.03 &   10.77$\pm$0.02 &   10.22$\pm$0.02  &   9.87$\pm$0.02  &   9.00$\pm$0.03  &  8.15$\pm$0.27  &      0.00      &   -1.91   \\ [0.3pt]   
 64 &  J061803.57+232837.9 & 16.52$\pm$0.19  &   15.54$\pm$0.18 &   14.99$\pm$0.14 &   13.62$\pm$0.03  &  13.36$\pm$0.04  &  10.11$\pm$0.07  &  7.47$\pm$0.12  &      2.97      &   -0.30   \\ [0.3pt]   
 65 &  J061749.30+232018.8 & 17.04$\pm$0.18  &   15.62$\pm$0.13 &   14.91$\pm$0.09 &   14.15$\pm$0.04  &  13.64$\pm$0.05  &  10.79$\pm$0.22  &  7.88$\pm$0.19  &      7.50      &   -0.77   \\ [0.3pt]   
 66 &  J061742.48+232427.5 & 17.02$\pm$0.19  &   15.52$\pm$0.12 &   15.06$\pm$0.09 &   14.35$\pm$0.04  &  13.75$\pm$0.06  &  11.39$\pm$0.30  &  7.94$\pm$0.00  &      8.45      &   -1.05   \\ [0.3pt]   
 67 &  J061615.59+222447.8 & 15.05$\pm$0.05  &   14.34$\pm$0.06 &   14.09$\pm$0.06 &   13.67$\pm$0.03  &  13.23$\pm$0.04  &  11.47$\pm$0.24  &  8.71$\pm$0.00  &      0.50      &   -1.41   \\ [0.3pt]   
 68 &  J061729.18+222152.8 & 14.85$\pm$0.03  &   13.85$\pm$0.04 &   13.31$\pm$0.03 &   12.56$\pm$0.04  &  12.03$\pm$0.04  &  10.26           &  8.17$\pm$0.28  &      3.60      & \nodata   \\ [0.3pt]   
 69 &  J061732.84+222308.1 & 13.59$\pm$0.03  &   13.07$\pm$0.03 &   12.80$\pm$0.03 &   12.54$\pm$0.06  &  12.11$\pm$0.10  &   9.66$\pm$0.27  &  7.87$\pm$0.24  &      0.00      &   -1.08   \\ [0.3pt]   
 70 &  J061527.24+225413.5 & 15.62$\pm$0.06  &   14.63$\pm$0.06 &   14.02$\pm$0.05 &   13.40$\pm$0.03  &  13.03$\pm$0.03  &   9.81$\pm$0.07  &  7.29$\pm$0.11  &      1.53      &   -0.55   \\ [0.3pt]   
 71 &  J061705.63+221812.7 & 14.98$\pm$0.04  &   14.41$\pm$0.06 &   14.09$\pm$0.05 &   13.69$\pm$0.03  &  13.19$\pm$0.04  &  11.48$\pm$0.29  &  8.17$\pm$0.27  &      0.00      &   -1.40   \\ [0.3pt]   
 72 &  J061637.15+224253.3 & 16.76$\pm$0.16  &   15.39$\pm$0.12 &   14.74$\pm$0.08 &   13.63$\pm$0.03  &  13.13$\pm$0.04  &  11.74$\pm$0.31  &  8.40$\pm$0.00  &      6.83      &   -1.46   \\ [0.3pt]   
 73 &  J061939.62+223711.9 & 16.69$\pm$0.17  &   15.52$\pm$0.14 &   14.72$\pm$0.09 &   13.84$\pm$0.03  &  13.52$\pm$0.04  &  11.29$\pm$0.29  &  8.18$\pm$0.26  &      2.69      &   -1.07   \\ [0.3pt]   
 74 &  J061711.32+224649.4 & 15.80$\pm$0.07  &   14.71$\pm$0.06 &   14.21$\pm$0.05 &   13.50$\pm$0.03  &  13.01$\pm$0.04  &  11.29$\pm$0.27  &  8.49$\pm$0.34  &      4.25      &   -1.38   \\ [0.3pt]   
 75 &  J061727.60+230007.0 & 16.18$\pm$0.08  &   15.16$\pm$0.09 &   14.24$\pm$0.06 &   13.58$\pm$0.03  &  12.89$\pm$0.03  &  10.99$\pm$0.21  &  8.55$\pm$0.00  &      0.00      &   -1.06   \\ [0.3pt]   
 76 &  J061810.85+230012.3 & 15.56$\pm$0.06  &   14.50$\pm$0.05 &   13.89$\pm$0.05 &   13.22$\pm$0.03  &  12.77$\pm$0.03  &  10.93$\pm$0.20  &  8.67$\pm$0.40  &      2.65      &   -1.31   \\ [0.3pt]   
 77 &  J061754.17+224851.2 & 15.06$\pm$0.04  &   14.35$\pm$0.05 &   14.15$\pm$0.05 &   13.75$\pm$0.04  &  13.34$\pm$0.07  &  10.87$\pm$0.50  &  7.33$\pm$0.29  &      0.28      &   -1.02   \\ [0.3pt]   
 78 &  J061720.91+225236.4 & 14.75$\pm$0.03  &   14.21$\pm$0.05 &   14.08$\pm$0.05 &   13.77$\pm$0.03  &  13.34$\pm$0.05  &  10.15$\pm$0.12  &  6.86$\pm$0.14  &      0.00      &   -0.62   \\ [0.3pt]   
 79 &  J061754.52+223312.0 & 14.52$\pm$0.04  &   14.02$\pm$0.04 &   13.93$\pm$0.05 &   13.68$\pm$0.03  &  13.25$\pm$0.05  &  11.19$\pm$0.21  &  8.39$\pm$0.00  &      0.00      &   -1.31   \\ [0.3pt]   
 80 &  J061639.71+222409.3 & 14.55$\pm$0.04  &   14.11$\pm$0.05 &   13.94$\pm$0.05 &   13.61$\pm$0.04  &  12.78$\pm$0.06  &  10.52           &  8.14$\pm$0.28  &      0.00      &  \nodata  \\ [0.3pt]   
 81 &  J061718.41+222210.4 & 14.42$\pm$0.04  &   13.85$\pm$0.06 &   13.59$\pm$0.05 &   12.97$\pm$0.03  &  12.41$\pm$0.04  &  12.06           &  8.59$\pm$0.00  &      0.00      &  \nodata  \\ [0.3pt]   
 82 &  J061756.26+224600.2 & 13.94$\pm$0.03  &   13.61$\pm$0.04 &   13.46$\pm$0.04 &   13.05$\pm$0.03  &  12.41$\pm$0.03  &  10.92$\pm$0.22  &  7.28$\pm$0.16  &      0.00      &   -1.44   \\ [0.3pt]   
 83 &  J061752.80+224957.8 & 14.99$\pm$0.07  &   14.25$\pm$0.07 &   13.92$\pm$0.05 &   13.41$\pm$0.04  &  13.00$\pm$0.04  &   9.82$\pm$0.09  &  6.99$\pm$0.15  &      1.20      &   -0.58   \\ [0.3pt]   
 84 &  J061715.37+223141.3 & 16.41$\pm$0.13  &   15.35$\pm$0.11 &   14.74$\pm$0.09 &   13.93$\pm$0.03  &  13.60$\pm$0.06  &  11.64$\pm$0.26  &  8.55$\pm$0.00  &      2.67      &   -1.25   \\ [0.3pt]   
 85 &  J061624.16+223046.2 & 16.91$\pm$0.19  &   15.69$\pm$0.16 &   15.40$\pm$0.18 &   14.07$\pm$0.04  &  13.79$\pm$0.07  &   9.68$\pm$0.07  &  7.32$\pm$0.12  &      5.85      &    0.12   \\ [0.3pt]   
 86 &  J061736.65+222450.8 & 14.53$\pm$0.03  &   13.94$\pm$0.04 &   13.73$\pm$0.04 &   13.40$\pm$0.03  &  13.00$\pm$0.06  &  11.59$\pm$0.51  &  8.79$\pm$0.00  &      0.00      &   -1.66   \\ [0.3pt]   
 87 &  J061823.18+224151.9 & 14.95$\pm$0.05  &   14.11$\pm$0.05 &   13.85$\pm$0.05 &   13.37$\pm$0.03  &  12.93$\pm$0.04  &  11.79$\pm$0.54  &  7.96$\pm$0.40  &      1.40      &   -1.77   \\ [0.3pt]   
 88 &  J061818.14+223451.4 & 14.12$\pm$0.04  &   13.74$\pm$0.05 &   13.53$\pm$0.05 &   13.08$\pm$0.06  &  12.28$\pm$0.05  &  10.43$\pm$0.32  &  7.68$\pm$0.51  &      0.00      &   -1.12   \\ [0.3pt]   
 89 &  J061809.01+223712.1 & 14.30$\pm$0.04  &   13.60$\pm$0.05 &   13.40$\pm$0.04 &   13.15$\pm$0.06  &  12.71$\pm$0.10  &  10.25           &  7.86$\pm$0.26  &      0.16      &  \nodata  \\ [0.3pt]   
 90 &  J061703.34+223629.0 & 14.19$\pm$0.03  &   13.72$\pm$0.04 &   13.73$\pm$0.04 &   13.39$\pm$0.03  &  12.98$\pm$0.04  &  10.92$\pm$0.20  &  8.61$\pm$0.00  &      0.00      &   -1.28   \\ [0.3pt]   
 91 &  J061719.90+225141.2 & 17.25$\pm$0.23  &   15.81$\pm$0.15 &   15.17$\pm$0.11 &   14.30$\pm$0.05  &  13.46$\pm$0.06  &   9.68$\pm$0.06  &  6.68$\pm$0.08  &      7.16      &    0.03   \\ [0.3pt]   
 92 &  J061633.37+230237.0 & 16.64$\pm$0.13  &   15.14$\pm$0.09 &   14.60$\pm$0.07 &   14.00$\pm$0.03  &  13.30$\pm$0.04  &  11.31$\pm$0.25  &  8.40$\pm$0.00  &      8.43      &   -1.25   \\ [0.3pt]   
 93 &  J061725.28+225210.8 & 14.62$\pm$0.03  &   13.97$\pm$0.05 &   13.78$\pm$0.04 &   13.63$\pm$0.04  &  13.12$\pm$0.05  &  10.00$\pm$0.15  &  6.99$\pm$0.23  &      0.00      &   -0.68   \\ [0.3pt]   
 94 &  J061954.07+222403.2 & 16.83$\pm$0.17  &   15.57$\pm$0.15 &   15.03$\pm$0.13 &   14.18$\pm$0.03  &  13.64$\pm$0.04  &   9.94$\pm$0.07  &  8.02$\pm$0.22  &      5.45      &   -0.16   \\ [0.3pt]   
 95 &  J061656.21+225005.6 & 16.65$\pm$0.14  &   15.43$\pm$0.12 &   14.92$\pm$0.09 &   13.91$\pm$0.03  &  13.31$\pm$0.04  &  11.75           &  8.95$\pm$0.00  &      5.03      &  \nodata  \\ [0.3pt]   
 96 &  J061807.51+220431.3 & 16.62$\pm$0.17  &   15.57$\pm$0.16 &   14.81$\pm$0.12 &   13.68$\pm$0.03  &  13.44$\pm$0.04  &   9.64$\pm$0.04  &  7.87$\pm$0.19  &      1.27      &   -0.05   \\ [0.3pt]   
 97 &  J061918.96+222627.3 & 16.58$\pm$0.15  &   15.52$\pm$0.14 &   14.82$\pm$0.10 &   13.63$\pm$0.03  &  13.30$\pm$0.04  &   9.37$\pm$0.04  &  7.13$\pm$0.11  &      1.89      &    0.09   \\ [0.3pt]   
 98 &  J061728.02+222153.6 & 14.86$\pm$0.05  &   13.87$\pm$0.05 &   13.38$\pm$0.03 &   12.75$\pm$0.04  &  12.21$\pm$0.04  &  10.29$\pm$0.46  &  7.63$\pm$0.19  &      3.64      &   -1.25   \\ [0.3pt]      
\end{longtable*}

\begin{figure*}
\centerline{\includegraphics[scale=0.3,angle=0]{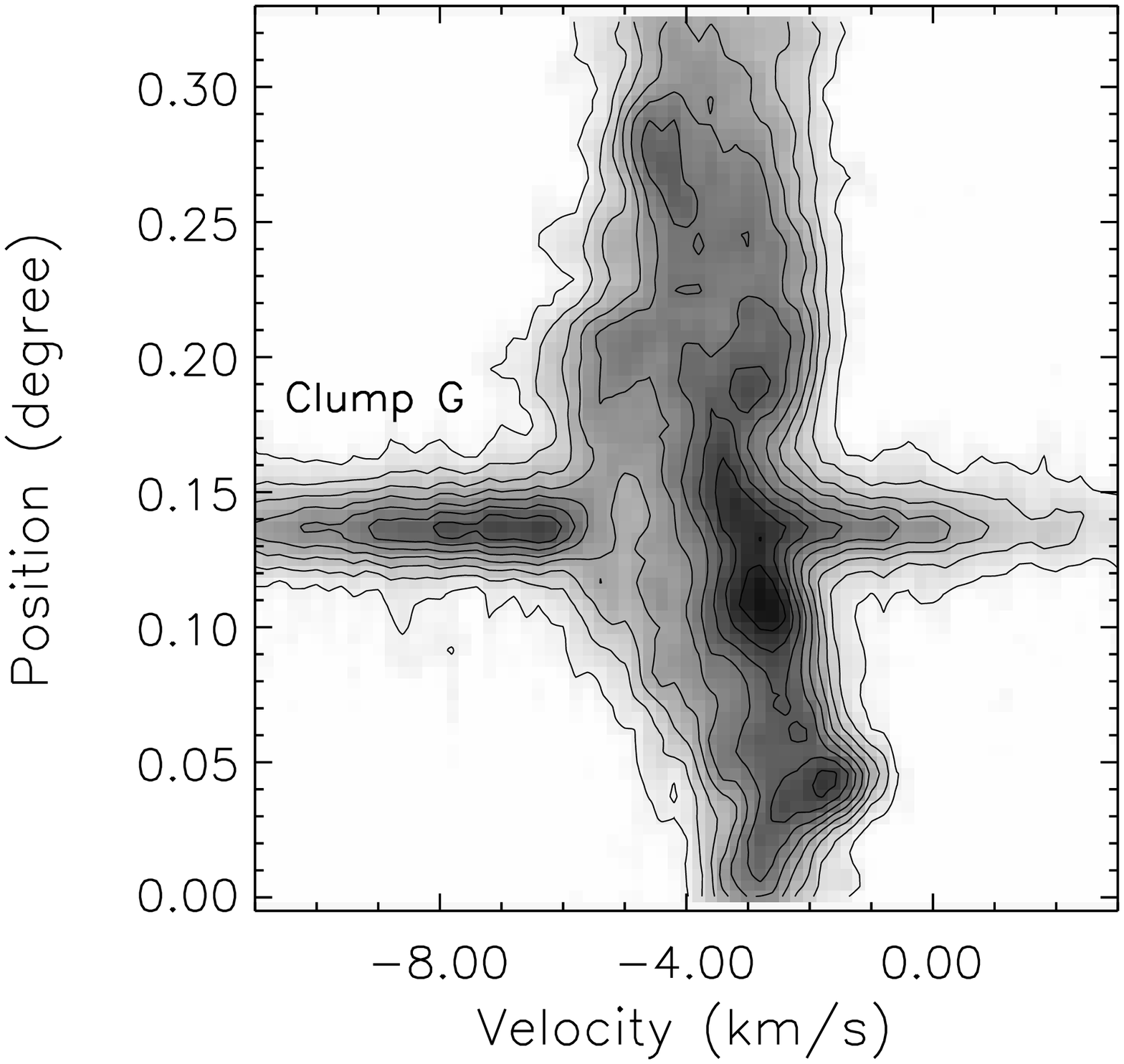}
            \includegraphics[scale=0.3,angle=0]{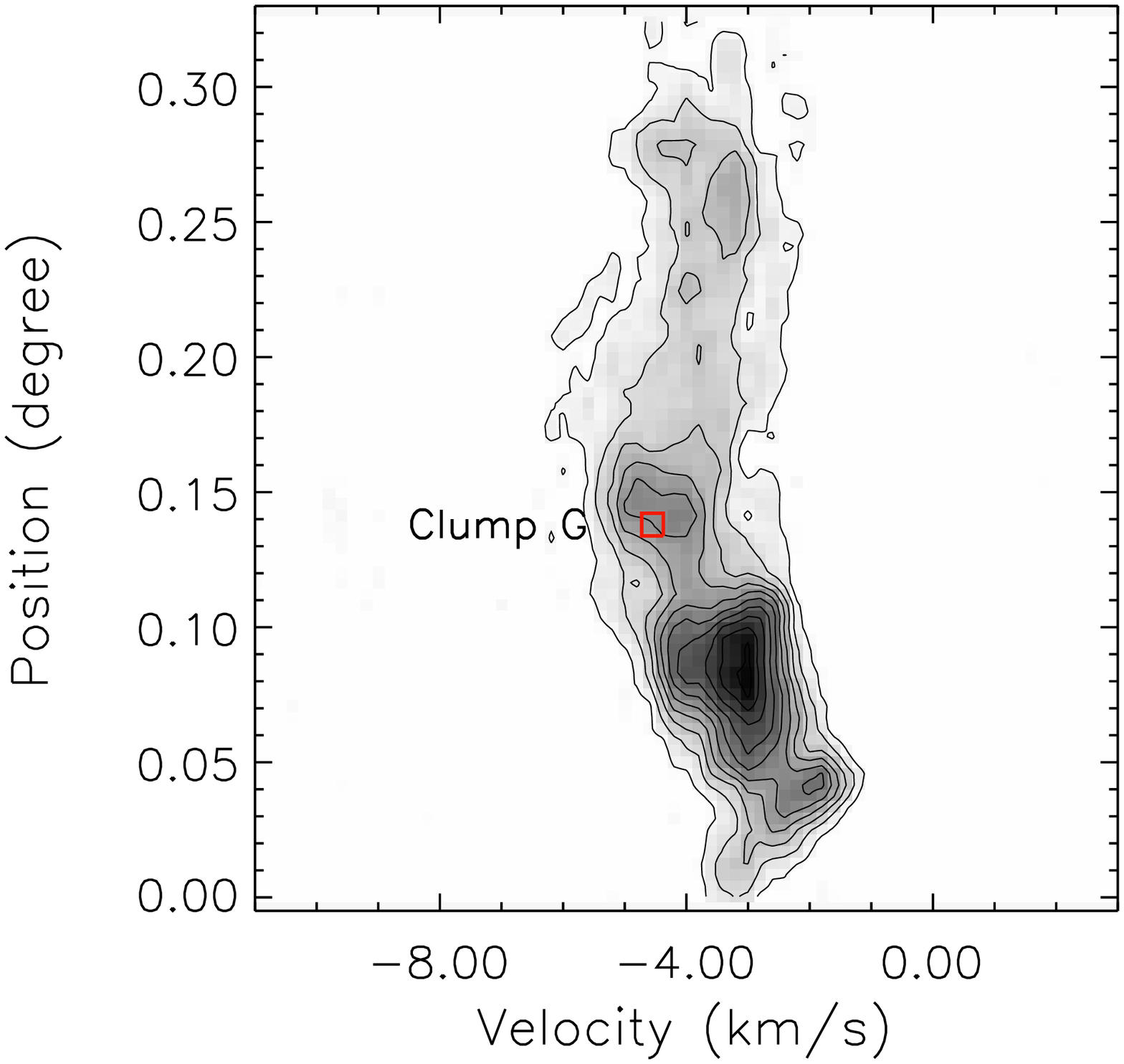}}
	    \caption{ Left: position--velocity diagram of the \twCO\ emission across
	    the shocked clump G along the southwest--northeast
	    (see the short arrow in Figure~3). Right: the \thCO\ emission along the
	    southwest--northeast. Clump G is labeled in the map.
	    The red box indicates the position of the 1720~MHz OH maser emission.
	    \label{fig12}}
	    \end{figure*}

\begin{figure*}
\centerline{\includegraphics[scale=0.3,angle=0]{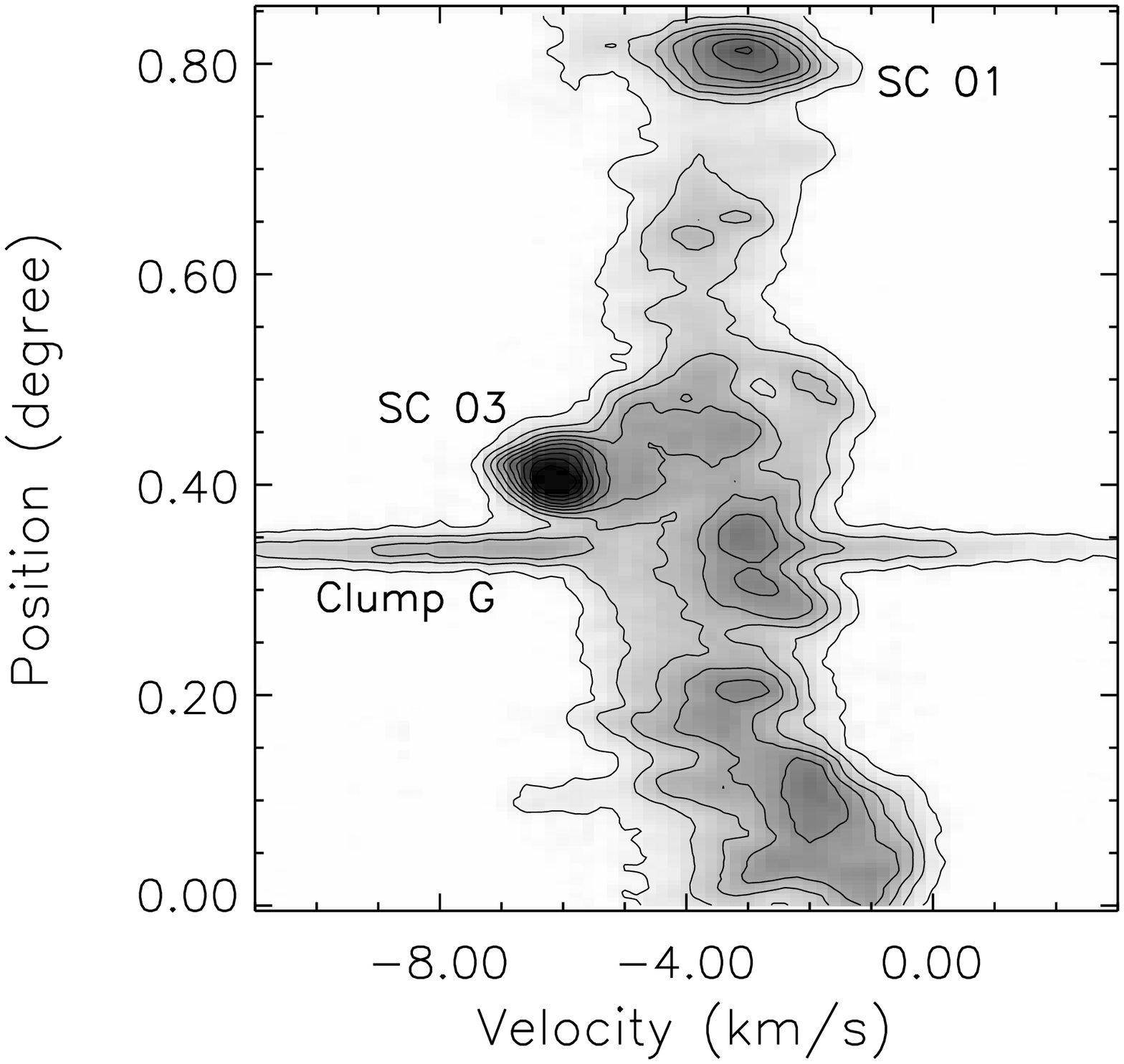}
	    \includegraphics[scale=0.3,angle=0]{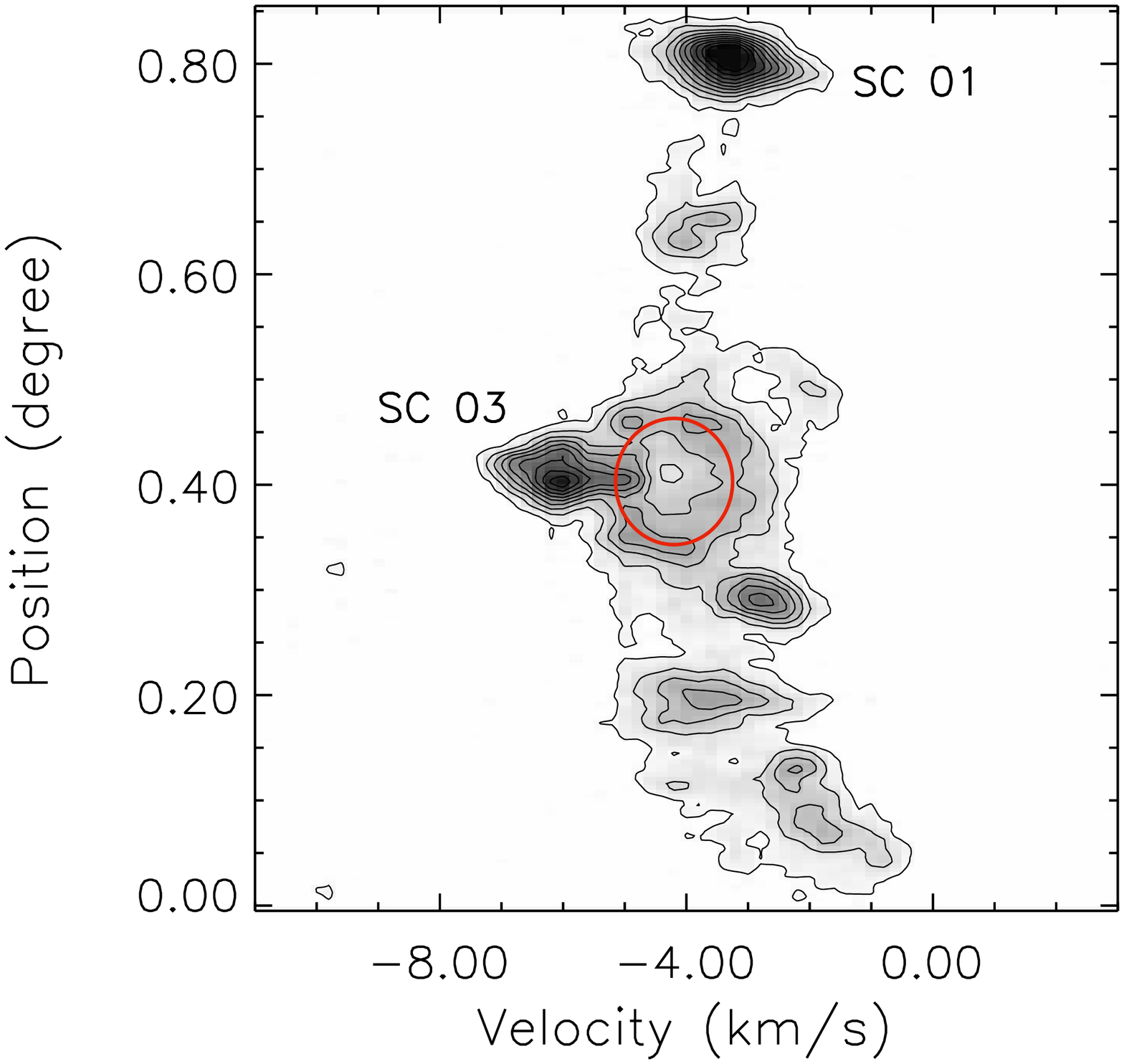}}
	    \caption{ Left: position--velocity diagram of the \twCO\ emission along the
	    southeast--northwest (see the long arrow in Figure~3). Clumps G,
	    SC~03, and SC~01 are labeled in the diagram. Right: the \thCO\ emission
	    along the southeast--northwest. The red ellipse indicates a
	    ring structure with a radius of 3\farcm6.
	    \label{fig13}}
\end{figure*}

The additional half-shell structure with the velocity range of 
$-5$~km~s$^{-1}$ to $-2$~km~s$^{-1}$ is also found to surround the 
northeastern half of the remnant (see Section 4.2). We suggest 
that the $-4$~km~s$^{-1}$ MCs, which are located majorly on the front of the
remnant, might be close to and physically related to SNR IC~443
combining the spatial morphology of the MCs and its velocity coincidence 
with the shocked clumps (see Table~1). 
The suggestion is supported by the study of the radio spectral index
distribution (see Figure~11 in Castelletti et al. 2011). In their study,
the coincidence between the distribution of the flat radio spectrum and that of
the high-density gas traced by CO in the interior of SNR IC~443 indicates
that the strong shocks are interacting with dense ambient medium
(Castelletti et al. 2011). Furthermore, the good morphological correlation 
between the
TeV emission and the distribution of the $-4$~km~s$^{-1}$ MCs
shows the possible connection between them (see Figure~12 in
Castelletti et al. 2011).

The LSR velocities of the preshock gas for shocked clumps
SC~05 and D seem to slightly deviate from $-4$~km~s$^{-1}$.
We consider that the velocity deviation of the two shocked clumps in
the southeast of the remnant is due to the turbulence in the molecular gas.
The velocity deviation between the two shocked clumps and the
$-4$~km~s$^{-1}$ MCs can be a result of the velocity gradient within the MCs.
Moreover, the \twCO\ PV diagram (Figure~17 in Lee et al. 2012) actually shows
the velocity gradient along the southeast--northwest. 
On the other hand, the shocked clumps E, F, and G 
have a similar LSR velocity at about $-4$~km~s$^{-1}$ and positive-velocity 
broad wing profiles, possibly due to the
shock moving nearly transverse to our line of sight. It is not contradictory 
that most of the $-4$~km~s$^{-1}$ MCs are located on the front side of 
SNR IC~443. 

To investigate the relationship between the shocked clump G
and SCs, we made the PV diagrams along the southeast--northwest
(the long arrow in the right panel of Figure~3). The arrow passes through the
main body of the $-4$~km~s$^{-1}$ MCs, and the PV diagrams are presented in 
Figure~13.
The prominent broad-line wing of the \twCO\ emission is from the
shocked clump G (the left panel in Figure~13). The SCs 03 and 01 are both 
discerned in the PV diagrams of the \twCO\ and \thCO\ emission.
The most attractive feature in the PV diagram of the \thCO\ emission
is that a ring structure of molecular gas (the red ellipse in Figure~13) 
overlaps with SC~03. In the right panel of Figure~13, the molecular ring 
structure with radius of 
3\farcm6 is visible in the velocity range of $-5$~km~s$^{-1}$ to 
$-3$~km~s$^{-1}$.
The $-6.2$~km~s$^{-1}$ SC~03 shows a filamentary structure ($\sim5'\times2'$;
see Figure~12 in Lee et al. 2012) 
and just superposes on the edge-on molecular ring structure.
It is unlikely that the filamentary SC~03 and the molecular ring
structure overlap by chance at different velocities along the line of 
sight. The SC~03 is possibly in physical contact with the $-4$~km~s$^{-1}$ MCs. 
Details of this ring feature will be investigated 
in a future paper. On the other hand the LSR velocity of SC~01
($\sim-3.3$~km~s$^{-1}$) is in the range of the $-4$~km~s$^{-1}$ MCs 
(Figures~9 and 11).
We suggest that the SCs seem to likely have a relation with the 
$-4$~km~s$^{-1}$ MCs.

\subsection{The Half-ring Structure}
We have seen that the partial shell structures of \twCO\
emission in the northeast are concentric with the bright radio border
of IC~443 and appear to confine the remnant's faint radio halo in the east.
As mentioned in Section 3.1.2, we can see that the northern and northeastern 
partial shells, together with the eastern small clumps, form a half-ring 
structure to surround SNR IC~443 (the left panel in Figure~7).
The LSR velocity range of the half-ring structure is
consistent with the $-4$~km~s$^{-1}$ CO component. This is also
consistent with the LSR velocity of the quiescent, ambient molecular
gas of the shocked clumps B, E, F, and G (Table~1).
The half-ring structure distribution along the northeastern half of
the remnant indicates the high density there.
On the contrary, the density is very low in the southwestern half of the 
remnant, in which SNR IC~443 displays the breakout structure. 
Moreover, the half-ring structure seems to confine the remnant's faint
radio halo in the northeastern part (Figure~7).

The LSR velocity consistency between the half-ring structure and
the $-4$~km~s$^{-1}$ MCs (Figure~3) and the positional coincidence between
the ring structure and the remnant's radio halo (Figure~7) indicate that 
the half-ring structure with an LSR velocity of $-4$~km~s$^{-1}$ is 
associated with SNR IC~443. We also note that there are no emissions in 
the other bands corresponding to the half-ring structure, and the structure 
is unlikely shocked by the SNR. The half-ring
structure is probably related to the stellar winds of the
remnant's massive progenitor star (see Section 4.3).

\subsection{SNR Physics of IC~443}
We have shown that SNR IC~443 is located in the environment of the
dense gas (Troja et al. 2008). The remnant was likely from 
a core-collapse explosion of a massive progenitor star,
although the tail of the pulsar wind nebulae (PWNe) does not point toward the 
geometric center of the remnant (Olbert et al. 2001).

The scenario that SNR IC~443 is evolving inside the preexisting
wind-blown bubble was suggested by several authors (Braun \& Strom
1986; Troja et al. 2006, 2008; Lee et al. 2008; Yamaguchi et al. 2009).
The half-ring structure (e.g., the left panel in Figure~7), which
is discovered in our \twCO\ and \thCO\
observations, appears to confine the faint radio halo.
We propose that the half-ring structure originates from the
interaction between the IC~443's progenitor winds and its surrounding MCs.
Assuming that the northeastern partial shell structures were formed by the gas 
swept by the wind of the remnant's massive progenitor, the volume ratio 
between the partial shell 
(a cylinder-like volume with a diameter of 1\farcm5 and a height of 7$'$
assumed) and the original rectangular pyramid (a volume with a base area of
$7'\times1\farcm5$ and a height of 26$'$ assumed) is about 0.14. 
Adopting the mean molecular density of the partial shell structures 
of $\sim10^3$~${\rm cm}^{-3}$ (see Section 3.1.2), the mean density 
in the original rectangular pyramid would be
$n_m\sim$140~${\rm cm}^{-3}$ for the cloudy ISM environment.
We find that the mean density of our estimation is consistent with the 
170~${\rm cm}^{-3}$ estimation of Braun \& Strom (1986).
The massive stellar winds will sweep a dense shell from the
homogenized region and leave a low-density bubble around the star.

Using formulae (51) and (52) in Weaver et al. (1977) and assuming
$n_m\sim$140~${\rm cm}^{-3}$ and $R_2\sim11.3$~pc (26$'$ at a distance of 
1.5~kpc), we compute the value of the mechanical
luminosity $L_w\sim0.6\times10^{34}\,{V_2^{3}}\,{\rm erg}\,{\rm s}^{-1}$,
which roughly corresponds to a value typical for the B0 star (Table~1 in Chen 
et al. 2013) for shaping a bubble with the expansion velocity of the order
of magnitude of 1~km~s$^{-1}$ (Table~2). 
Since the expanding velocity of the half-ring structure is very low,
we think that the bubble is nearly stalled in the ISM.
Therefore, we use the linear relationship to assess the initial 
masses of the progenitors for SNRs associated with MCs
(Chen et al. 2013). Considering the blowout structure
of SNR IC~443, the remnant's progenitor mass of 15--19~$M_{\odot}$
is a lower limit, which is consistent with the above 
calculation for a B0 star.

\subsection{Star Formation in the Region of SNR IC~443}
Generally, Class I objects are likely younger on average than
Class II YSOs. In Figure~11 (see Section 3.2), we find that the
Class I and II YSOs are much more projectively concentrated in the region of
the northeastern part of SNR IC~443. Interestingly,
most of Class I YSOs (blue triangles) are located along
the boundary of the bright radio shell of the remnant.
Because of the coincidence between the distribution of our
identified YSO candidates and the remnant's morphology, it is
unlikely that these YSO candidates are the foreground and background
stars. Therefore we suggest that these YSOs projected on
the FOV of the remnant are probably associated with SNR IC~443.
Considering the age difference between the YSOs ($10^{5}$--$10^{7}$~yr)
and the remnant ($10^{3}$--$10^{4}$~yr), we suggest 
that these YSOs are likely to be triggered by the stellar winds
from the progenitor of IC~443 (see Section 4.3).

\begin{figure*}
	    \includegraphics[scale=1.0,angle=0]{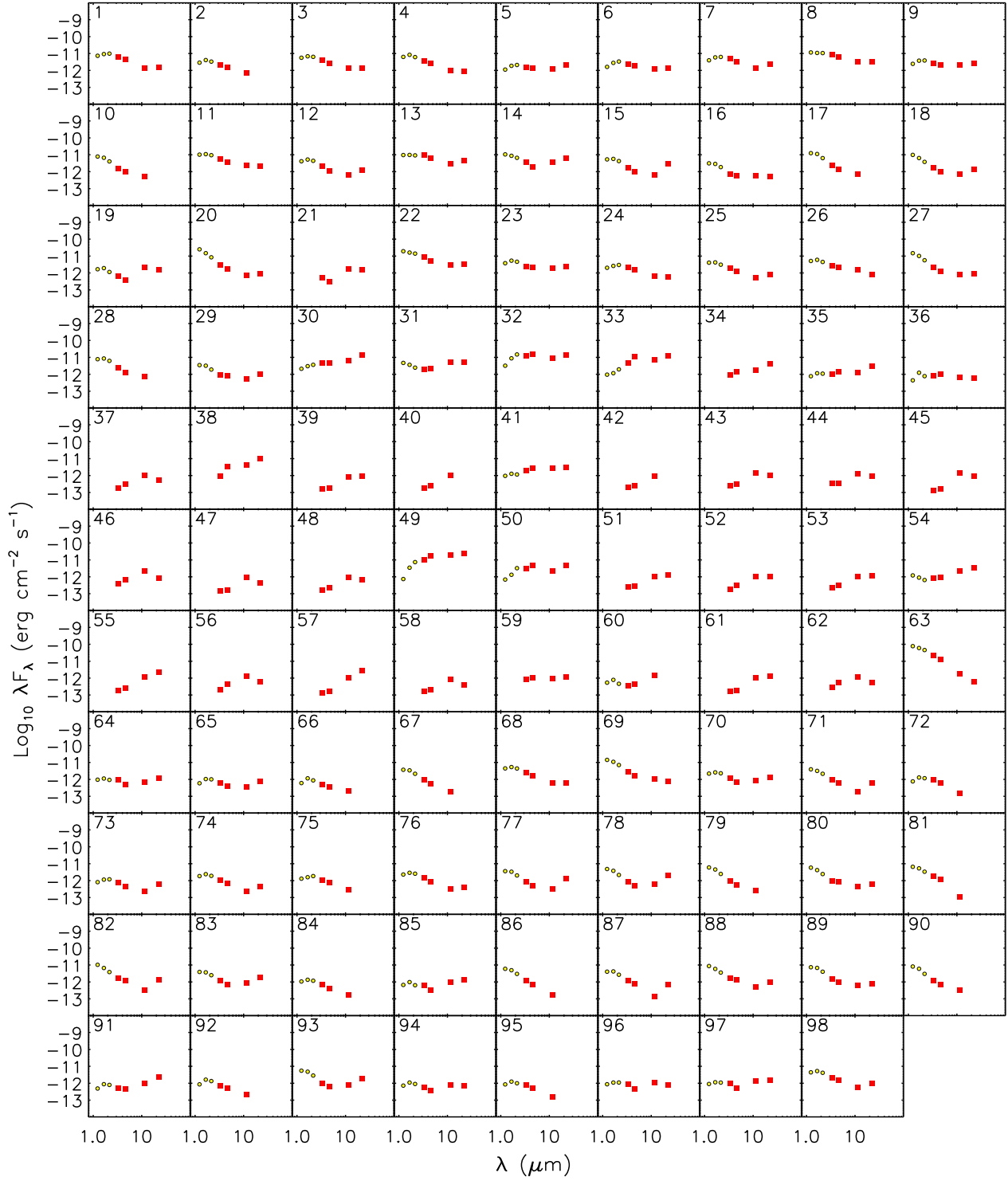}
	    \caption{SEDs for the 98 YSO candidates. Yellow points are 2MASS data and
	    red squares are $WISE$ data. Each plot has an ID, which
	    can be found in Table~3.
	    \label{fig14}}
\end{figure*}

Xu et al. (2011) used the 2MASS All-Sky Point Source database in the
near-IR $J$(1.25 $\mu$m), $H$(1.65 $\mu$m), and $K_{\rm S}$ (2.17 $\mu$m)
bands to look for primary tracers of star formation activity in the
vicinity of IC~443. They identified 1666 YSO candidates in a search
circle around IC~443 within a 25$'$ radius, whose number is greater than
ours. 
It is probable that some YSO candidates selected by Xu et al. (2011) 
have flat SEDs and should be rejected (e.g., Rebull et al. 2011). 
We show the SED diagrams for the 98 sources in Figure~14.
Actually, using 2MASS data, we identified 260 YSOs (including
classical T Tauri stars and Herbig Ae/Be stars) with an S/N greater 
than 3. Xu et al. (2011)
proposed that the distribution of YSO candidates may show a shell
morphology (see Figure~5(b) in their paper). Interestingly, the
similar shell structure of the distribution of 
our YSO candidates can be seen roughly in Figure~11.
The radius of the shell structure is $\sim$20$'$, which is surrounded
by the radio halo of SNR IC~443.

Whitworth et al. (1994) have studied the fragmentation of the 
shocked dense shells swept up by an expanding stellar wind
bubble. Adopting a reasonable value of $n_0=n_m\sim$140~${\rm cm}^{-3}$,
$L_w\sim1\times10^{34}\,{\rm erg}\,{\rm s}^{-1}$ (see Section 4.3),
and the effective sound speed $a_s$=0.2~km~s$^{-1}$, we can estimate the 
time at which the fragmentation begins, $t_{\rm fragment}\sim$5.7~Myr and the
radius of the shell at that time, $R_{\rm fragment}\sim$10.8~pc.
This result is roughly compatible with the value of 9~pc (20$'$ at 1.5~kpc).
We fit YSO models to 75/98 sources combined 2MASS+$WISE$ bands (see
Columns 3--8 in Table~3) based on the work of Robitaille et al. (2007); 57/75
sources were well fitted ($\chi^2$ less than 12). The estimated age is between
0.01~Myr and several megayears. We emphasized that the SED models are 
often highly
ambiguous because the models show a high degree of degeneracy. Therefore, the
stellar parameters are often uncertain, especially if only IR photometry is
available. Spectroscopic information is needed to estimate the actual age of
these YSO candidates.
On the other hand, since these sources are Class I and 
Class II sources, the typical lifetimes for these sources are considered 
to be 0.5~Myr and 2--3~Myr (Evans et al. 2009), respectively.
It is also compatible with the general evolutionary scheme of
the progenitor's wind of SNR IC~443. 

It is common that many low-mass stars will experience nearby supernovae.
When a massive star dies as a supernova, the ejecta of it will
expand freely through the hot, tenuous interior of an \mbox{H\,\textsc{ii}} 
region. Subsequently, the gaseous ejecta from the supernova hits 
a protoplanetary disk to form a bow shock structure. 
Chevalier (2000) showed that a protoplanetary disk around a young 
star can survive from the nearby supernova explosion.
Our solar system had very likely experienced such a process and survived 
from the extreme environment. 
Recent isotopic analysis of meteorites have revealed that the Sun
and the solar system formed near a massive star and experienced 
a supernova explosion (Hester \& Desch 2005).
Briefly, the explosion of the supernova plays an essential role for 
the formation and evolution of the solar system (Tachibana et al. 2006).  

\subsection{The High-energy Emission of SNR IC~443}
For SNR IC~443, the origin of the $\gamma$-ray emission is
suggested to be from hadronic collisions. Abdo et al. (2010)
summarized the locations and extensions of the $\gamma$-ray emission
from IC~443 detected by different telescopes (e.g., EGRET: Hartman
et al. 1999; MAGIC: Albert et al. 2007; VERITAS: Acciari et al.
2009; and $Fermi$ LAT: Abdo et al. 2010). Based on these observations,
Torres et al. (2010) used two MC models to explain the GeV--TeV
connection in the vicinity of SNR IC~443. Li \& Chen (2012),
however, used the MC (with geometric volume) model to fit the
GeV--TeV spectra. In either case, the $-4$~km~s$^{-1}$ MC
component, for which the centroid 
of the molecular gas near the center of the remnant 
is consistent with the centroid of the GeV--TeV emission, could
play an essential role for the observed $\gamma$-ray emission.
Recently, Ackermann et al. (2013) identified the characteristic
pion-decay feature in the gamma-ray emission. It
provides direct evidence that cosmic-ray protons are
accelerated in SNR IC~443. The location of the gamma-ray emission
is close to the shocked clump G. For shocked clump G, the
$-4.4$~km~s$^{-1}$ systematic velocity of the quiescent gas can
be seen in Table~1, where the velocity is consistent with the result
of the 1720~MHz OH observation (clump G, $-4.55$~km~s$^{-1}$;
Hewitt et al. 2006).
Additionally, the remarkable morphological coincidence between the TeV
emission and the $-4$~km~s$^{-1}$ MC along the southeasta--northwest 
(see Figure~12(c) in Castelletti et al. 2011)
indicates the connection between them.
Therefore, the $-4$~km~s$^{-1}$ MC is adjacent to SNR IC~443.

\section{SUMMARY}
Detailed millimeter CO studies have been performed toward SNR
IC~443. By combining
other multiwavelength data, we summarize the main results of our
analysis as the follows.

1. The shocked clumps B--G (named from previous studies; e.g.,
Huang et al. 1986; Dickman et al. 1992), which appear
to be connected by faint diffuse emission,
are clearly identified in our new \twCO\ maps. The
spectra of these clumps are consistent with previous results.

2. The $-4$~km~s$^{-1}$ MC extending along the southeast--northwest
is mainly located on the front of SNR IC~443. It is close
to and associated with the remnant.

3. We find that a half-ring structure centered at
$(\RA{06}{16}{57}.1,\Dec{22}{37}{06})$ with a radius of 26$'$
appears to surround the northern part of SNR IC~443. We suggest
that this structure in the velocity range of $-5$~km~s$^{-1}$ to 
$-2$~km~s$^{-1}$
is associated with IC~443. It may not be interacted with
SNR IC~443 directly, but associated with the stellar winds of its
massive progenitor star.

4. The progenitor of the remnant is probably 
a 15--19~$M_{\odot}$ B0 star.

5. Sixty-two YSO candidates, mainly concentrated along the boundary of the
remnant's bright radio shell, are likely to be triggered by
the stellar winds from the massive progenitor of SNR IC~443.

\acknowledgments
The authors acknowledge the staff members of the Qinghai Radio
Observing Station at Delingha for their support in observation.
We thank the anonymous referees for valuable advices and comments.
The work also used the data from (1) the $WISE$, 
which is a joint project of the University of California, 
Los Angeles, and the Jet Propulsion Laboratory (JPL), California 
Institute of Technology (Caltech), funded by the National Aeronautics 
and Space Administration (NASA); (2) the Two Micron All Sky Survey, 
which is a joint project of the University of Massachusetts and the 
Infrared Processing and Analysis Center/California Institute of 
Technology, funded by NASA and the National Science Foundation; and
(3) the GLIMPSE data with the $Spitzer$ $Space$ $Telescope$, which is 
operated by the Jet Propulsion Laboratory, California Institute of 
Technology under a contract with NASA. We also thank Dr. Lee,~J.~J., 
for providing the radio continuum data.
This work is supported by NSFC grants 11103082, 11103088, 11233001, and
11233007. Y.S. acknowledges support from grant BK2011889, M.F. acknowledges
support by the NSFC through grants 11203081, and Y.C. acknowledges support
from the grant 20120091110048 by the Educational Ministry of China.
The work is a part of the Multi-Line Galactic Plane Survey in CO and its
Isotopic Transitions, also called the Milky Way Imaging Scroll
Painting, which is supported by the Strategic Priority Research Program
-- The Emergence of Cosmological Structures of the Chinese Academy of 
Sciences, Grant No. XDB09000000.

\end{document}